\documentclass[preprint,12pt]{elsarticle}

\let\today\relax
\makeatletter
\def\ps@pprintTitle{%
    \let\@oddhead\@empty
    \let\@evenhead\@empty
    \def\@oddfoot{\footnotesize\itshape
         {} \hfill\today}%
    \let\@evenfoot\@oddfoot
    }
\makeatother

\usepackage{lineno}

\usepackage{amsmath,amsthm,amssymb}
\usepackage{graphicx}
\usepackage[colorlinks=true, allcolors=blue]{hyperref}
\usepackage{float}

\newtheorem{theorem}{Theorem}
\newtheorem{lemma}{Lemma}
\newtheorem{corollary}{Corollary}

\usepackage[top=2.5cm,bottom=2.5cm,left=2.5cm,right=2.5cm]{geometry}

\journal{Arxiv}

\begin{document}

\begin{frontmatter}



\title{Nonlinear dynamics of CAR-T cell therapy}


\cortext[cor1]{fassoni@unifei.edu.br}
\author[unifei,imb]{Artur C. Fassoni\corref{cor1}} 
\author[unifei]{Denis C. Braga}

\affiliation[unifei]{organization={Instituto de Matemática e Computação, Universidade Federal de Itajubá},
            city={Itajubá},
            country={Brazil}}
            
\affiliation[imb]{organization={Institute for Medical Informatics and Biometry, Technische Universität Dresden},
            city={Dresden},
            country={Germany}}

\begin{abstract}
Chimeric antigen receptor T-cell (CAR-T) therapy is considered a promising cancer treatment. The dynamic response to this therapy can be broadly divided into a short-term phase, ranging from weeks to months, and a long-term phase, ranging from months to years. While the short-term response, encompassing the multiphasic kinetics of CAR-T cells, is better understood, the mechanisms underlying the outcomes of the long-term response, characterized by sustained remission, relapse, or disease progression, remain less understood due to limited clinical data. Here, we analyze the long-term dynamics of a previously validated mathematical model of CAR-T cell therapy. We perform a comprehensive stability and bifurcation analysis, examining model equilibria and their dynamics over the entire parameter space. Our results show that therapy failure results from a combination of insufficient CAR-T cell proliferation and increased tumor immunosuppression. By combining different techniques of nonlinear dynamics, we identify Hopf and Bogdanov-Takens bifurcations, which allow to elucidate the mechanisms behind oscillatory remissions and transitions to tumor escape. In particular, rapid expansion of CAR-T cells leads to oscillatory tumor control, while increased tumor immunosuppression destabilizes these oscillations, resulting in transient remissions followed by relapse. Our study highlights different mathematical tools to study nonlinear models and provides critical insights into the nonlinear dynamics of CAR-T therapy arising from the complex interplay between CAR-T cells and tumor cells.
\end{abstract}




\end{frontmatter}


\section{Introduction}

Chimeric antigen receptor T-cell (CAR-T) therapy represents a significant advance in cancer treatment, particularly for hematological malignancies such as acute leukemia and lymphomas. This approach involves extracting T cells from patients' blood, genetically modifying them to express chimeric antigen receptors (CARs), and reintroducing them into the patient's body. Engineered to target specific antigens expressed by tumor cells, CAR-T cells offer the potential to improve outcomes where conventional therapies have failed \cite{grupp2013chimeric}.

\begin{figure}
    \begin{center}
    \includegraphics[width=0.5\linewidth]{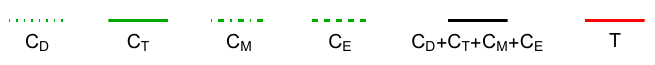}    
    \vspace{-0.5cm}
    
    \includegraphics[width=0.99\linewidth]{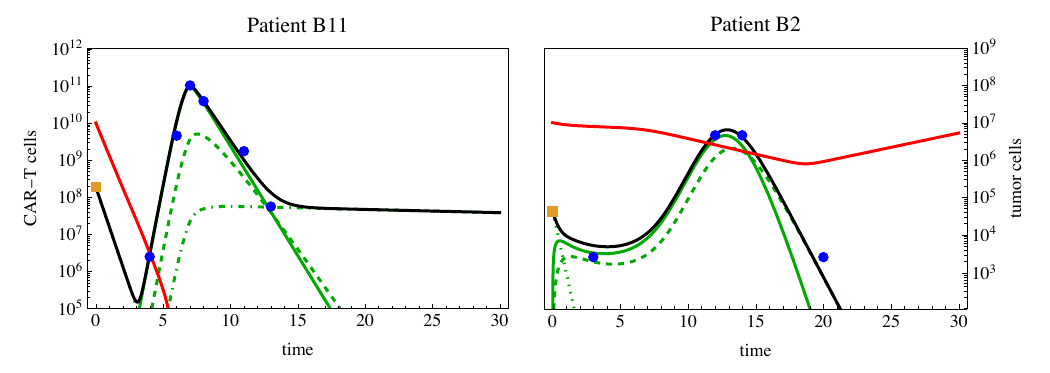}
    \end{center}    
    \caption{Fits of model \eqref{system1simp} to time courses of two patients illustrating the multiphasic dynamics of CAR-T cells and the different tumor  responses: remission (patient B11) and stable disease (patient B2). CAR-T cells are shown in green, tumor cells in red, and the total CAR-T cell population in black. Patient data are shown as blue dots. Data and parameter values obtained from \cite{paixao2022modeling}.}
    \label{fig:fits}
\end{figure}

The dynamical response of CAR-T therapy can be conceptually divided into two distinct stages: the short-term response, lasting approximately one or a few months, and the long-term treatment outcome spanning several months to years. The short-term response is characterized by multiphasic kinetics of CAR-T cells (Figure \ref{fig:fits}), arising from the interplay between different phenotypes and encompassing four phases \cite{stein2019tisagenlecleucel,liu2021model,paixao2022modeling}: a distribution phase marked by the decay of injected CAR-T cells and their migration to the tumor site; an expansion phase with significant proliferation of CAR-T cells triggered by antigen recognition; a contraction phase where CAR-T cell levels decline due to reduced tumor load; and a persistence phase dominated by memory CAR-T cells with slower decline attributed to their extended lifespan. Clinical documentation of this response is typically facilitated by regular blood sampling, often conducted during inpatient observation periods. Typically, the persistence phase commences one or a few months after CAR-T injection.

Conversely, the long-term response unfolds over an extended timeframe and lacks detailed time courses due to reduced blood sampling. This phase is characterized by sustained remission, molecular relapse detected through routine examinations, or disease progression marked by the return of symptoms, the latter two indicative of therapy failure. Common reasons for relapse include suboptimal CAR-T cytotoxic efficiency or expansion, tumor-mediated inhibition of CAR-T and immune cells, and the emergence of resistant tumor cells, notably those lacking target antigens \cite{santurio2024mechanisms,shah2019mechanisms}.

A previous publication by one of us presented the development of a differential equation model that effectively captures the multiphasic kinetics in the short-term response and provides mechanistic insights into how it emerges through the phenotypic differentiation of CAR-T cells \cite{paixao2022modeling}. Regarding the long-term response, model simulations with patient-specific parameter fits reproduced both treatment tumor control and escape (Figure \ref{fig:fits}), and also presented sustained oscillations in CAR-T cell-tumor kinetics, suggesting the existence of limit cycles. However, this previous work primarily focused on elucidating the short-term dynamics of CAR-T therapy, leaving the comprehensive understanding of the model's long-term behavior unresolved.

In this manuscript, we aim to dissect the long-term behavior of this previous CAR-T cell therapy model. While the original model comprises a nonautonomous system of five differential equations, the model behavior in the long-term is described by a reduced three-dimensional autonomous subsystem. Therefore, disregarding the transient terms, we conduct a comprehensive stability and bifurcation analysis of the three-dimensional permanent system to uncover the underlying mechanisms governing the long-term therapeutic response. Our study includes a complete examination of the model equilibria and their dynamics across the entire parameter space. This investigation provides a profound understanding of how the model parameters characterizing tumor and CAR-T cell attributes determine the long-term therapy response. Specifically, we demonstrate that therapy failure does not stem from a singular factor but rather from a confluence of inadequate CAR-T cell proliferation coupled with elevated tumor immunosuppression. Furthermore, we rigorously establish the existence of Hopf and Bogdanov-Takens bifurcations. Such a mathematical analysis translates into the therapeutic context and shows that oscillatory remissions occur when CAR-T cell expansion is rapid. Moreover, by increasing tumor immunosuppression, which leads to the collapse of a limit cycle with a homoclinic loop, these oscillations can be destabilized, leading to a state of transient or false remission followed by tumor escape. In summary, by exploring the consequences and ramifications of the model's nonlinear dynamics, we provide valuable insights into the determinants of treatment efficacy.

This manuscript is organized as follows. In Section \ref{sec:results}, we present the original model and its reduction to the permanent system. Then we outline our mathematical results in an intuitive, non-formal manner, illustrated with two-dimensional bifurcation diagrams, phase portraits, and numerical simulations accompanied by biological interpretation. The findings are presented with precise statements and rigorous proofs in Section \ref{sec:proofs}. Finally, we provide the context of previous work in this area and discuss our results in Section \ref{sec:conclusion}.

\section{Results}
\label{sec:results}

\subsection{Model for CAR-T cell therapy and reduction to permanent system}

Our previous model for the multiphasic kinetics of CAR-T cell therapy comprises four compartments representing CAR-T cell phenotypes, namely distributed, effector, memory, and exhausted CAR-T cells, denoted as $C_D$, $C_T$, $C_M$, and $C_E$, respectively, along with one compartment of antigen-expressing tumor cells, denoted as $T$. The model is given by the following system of differential equations:
    \begin{subequations}
        \begin{align}
            \dfrac{d{{C}_D}}{dt}&= -\beta C_D -\eta C_D
            \ ,\label{distsimp} \\
            \dfrac{dC_{T}}{dt} &= \eta {C}_{D} + \left(r_{{\min}}+\frac{p_1}{1+p_2 t^{p_3}} \right) \frac{ T }{A + T} C_{T} - (\xi + \epsilon + \Lambda) C_{T} + \theta T C_{M} - \alpha TC_{T},
            \label{ct1simp} \\
		    \dfrac{dC_M}{dt} &= \epsilon  {{C_T}}
		    - {\theta T C_{M} } - \mu C_{M}\ ,\label{cmsimp}  \\ 
		    \dfrac{dC_E}{dt} &= \Lambda C_T - \delta C_E \ ,
		    \label{cesimp}  \\
		    \dfrac{dT}{dt} &= rT(1-b T) - \gamma \frac{(C_D+C_T)T}{a + C_D+C_T+ dT }\ .\label{t}
        \end{align}
        \label{system1simp}
    \end{subequations}

The model assumes that initially injected CAR-T cells exhibit the distributed phenotype, transitioning thereafter to the effector phenotype at a rate $\eta$. Upon antigen recognition, effector cells expand with a time-dependent rate that commences at its maximum value, $r_{\min}+p_1$, decreasing to the basal expansion rate, $r_{\min}$. Effector cells also transition to the memory phenotype at a rate $\epsilon$, and become exhausted at a rate $\Lambda$. In the presence of tumor cells, memory cells revert to effector at a rate $\theta$. Distributed, effector, memory and exhausted CAR-T cells have death rates $\beta$, $\xi$, $\mu$, and $\delta$, respectively. Tumor cells present a logistic growth with rate $r$ and carrying capacity $b^{-1}$, and exert an immunosuppressive effect on effector CAR-T cells, modeled by the mass-action law with a constant $\alpha$. Both distributed and effector CAR-T cells impose cytotoxic effects on tumor cells at a rate $\gamma$, with the Holling-type-II-like functional response given in \eqref{t}, inspired by previous models by De Boer and De Pillis \cite{de2013quantifying,de2005validated}.

Figure \ref{fig:fits} presents examples fits of model \eqref{system1simp} obtained in \cite{paixao2022modeling}, to time courses of two patients showing different responses: remission (Patient B11) and stable disease (Patient B2). Such outcomes suggest the existence of different attractors for the solutions of \eqref{system1simp} and show that even an initial expansion of CAR-T cells and decrease in the tumor load may be followed by tumor regrowth or persistence. Therefore, a deeper understanding of the outcomes of CAR-T therapy as described by system \eqref{system1simp} depends on characterizing its attractors and bifurcation diagrams.

This task can be achieved by analyzing the reduced autonomous system which describes the permanent behavior of solutions of \eqref{system1simp}. Such reduction is based on the following remarks. First, since the solution of \eqref{distsimp} is $C_D(t)=C_D(0)e^{-(\beta+\eta)t}$, we have $C_D\to 0$ for $t \to \infty$. Second, the nonautonomous expansion rate in \eqref{ct1simp} converges to $r_{\min}$ as $t\to \infty$. Finally, equation \eqref{cesimp} is decoupled and may be disregarded. Thus, the long-term behavior of solutions of system \eqref{system1simp} is equivalent to solutions of the reduced system given by
\begin{subequations}
        \begin{align}
\dfrac{dC_T}{dt} &= r_{\min} \dfrac{ T }{A + T} C_T - (\xi + \epsilon + \Lambda) C_{T}+ \theta T C_{M} - \alpha TC_{T}  ,  \label{ct1simpred} \\
\dfrac{dC_M}{dt} &= \epsilon C_T - \theta T C_{M} - \mu C_{M},\label{cmred}  \\ 
\dfrac{dT}{dt} &= rT(1-b T) - \gamma \dfrac{C_T}{a +  C_T + d T  } T. \ \label{tred}
        \end{align}
\label{system1red}
\end{subequations}
\noindent We refer to this three-dimensional autonomous system as the \textit{permanent system} since it portrays the asymptotic behavior of the five-dimensional nonautonomous system \eqref{system1simp}.

\subsection{Summary of nonlinear dynamics}

Our main results are given by Theorems \ref{thmstb}, \ref{thmhop} and \ref{thmbt} presented in Section \ref{sec:proofs}. In the following, we present their contents in a informal way in order to provide a straightforward description of our results with their biological implications. 


To simplify the mathematical notation, we analyzed the dimensionless form of system \eqref{system1red} given by
    \begin{subequations}
        \begin{align}
            \dfrac{d x}{d \tau} &= p\dfrac{xz}{k+z}  - wx + q z y - uzx,\\
	        \dfrac{d y}{d \tau} &= s x - qzy- mq y, \\
	        \dfrac{d z}{d \tau} &= z(1-z) - v \dfrac{xz}{c + x+z},
        \end{align}
        \label{sys2}
    \end{subequations}
\noindent where $x = (b/d) C_T$, $y = (b/d) C_M$, and $z = bT$ represent densities of effector CAR-T, memory CAR-T and tumor cells, respectively, and
\begin{equation}
    \tau = rt,\
    p=\displaystyle\frac{r_{\min}}{r},\
    u=\displaystyle\frac{\alpha}{r b}, \
    v=\displaystyle\frac{\gamma}{r},\
    w=\displaystyle\frac{\xi+\epsilon+\Lambda}{r}, \
    m=\displaystyle\frac{\mu b}{\theta}, \
    k=bA,\
    q=\displaystyle\frac{\theta}{r b}, \
    s=\displaystyle\frac{\epsilon}{r}, \   
    c=\displaystyle\frac{ab}{d}.
    \label{nondimpars}
\end{equation}
All parameters are assumed to be positive. We also assume that $v>1$, otherwise the maximum tumor kill rate $\gamma$ would be less than the tumor growth rate $r$, leading to an ineffective treatment. Since $w>s$, the parameter space is defined as $$\Lambda=
\{\sigma=(p,u,v,w,m,k,q,s,c) \in \mathbb{R}^9\;|\;p,u,m,k,q,s,c>0,\;w>s,\;v>1\}.$$ 

In Theorem \ref{thmstb}, we show that system \eqref{sys2} has the following equilibria with the following stability properties:
\begin{enumerate}
    \item The trivial equilibrium $E_0=(0,0,0)$, which is always a saddle point, with a two-dimensional stable manifold given by the plane $z=0$.
    \item The tumor escape equilibrium $E_T=(0,0,1)$, which may be either a stable node or a saddle with a two-dimensional stable manifold.
    \item Two nontrivial equilibrium points, given by $E_C=(x_2,y_2,z_2)$ and $E_S=(x_3,y_3,z_3)$, satisfying $0<z_2<z_3<1$ whenever both are positive (i.e., $x_i,y_i,z_i> 0$). Since $z_2$ is small, $E_C$ is denoted as the tumor control equilibrium point, while $E_S$ is denoted as the separation point (see below). If $E_C$ ($E_S$) is positive, the Jacobian matrix of \eqref{sys2} evaluated at $E_C$ ($E_S$) has at least one real negative (positive) eigenvalue.
\end{enumerate}

\begin{figure}
    \begin{center}
    \begin{minipage}{0.8\linewidth}
    \textbf{\large{a}}\vspace{-0.1cm}\\
    \includegraphics[width=\linewidth]{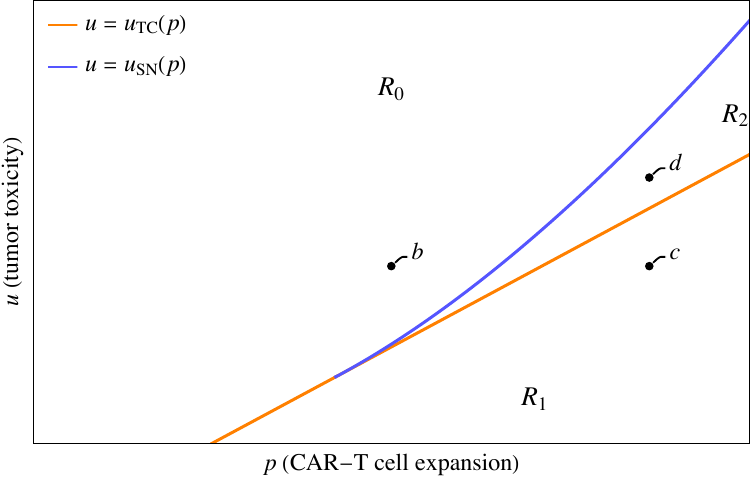}
    \end{minipage}
    \end{center}
    \vspace{0.4cm}
    
    \textbf{\large{b}} \hspace{4.5cm} {\textbf{\large{c}}} \hspace{4.5cm} {\textbf{\large{d}}} \vspace{-0.3cm}\\
    \includegraphics[width=0.32\linewidth]{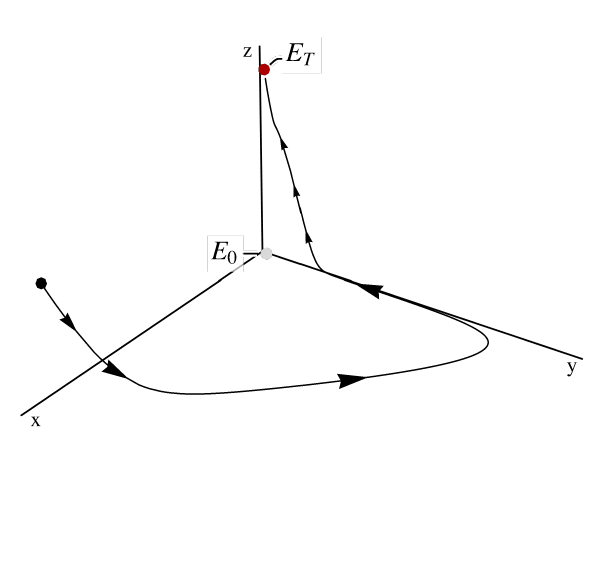}
    \includegraphics[width=0.32\linewidth]{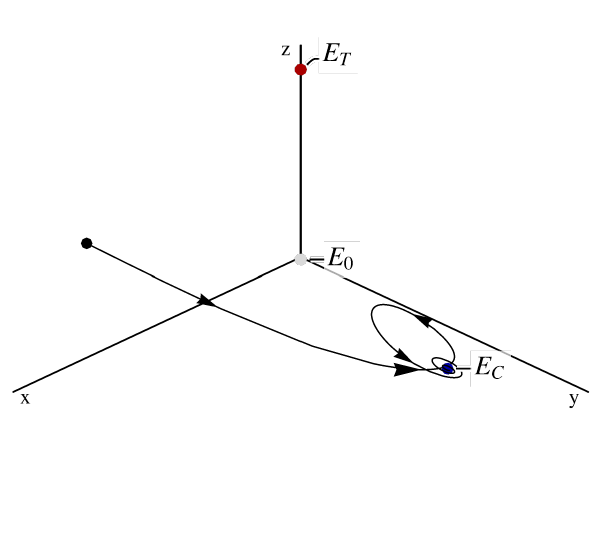}
    \includegraphics[width=0.32\linewidth]{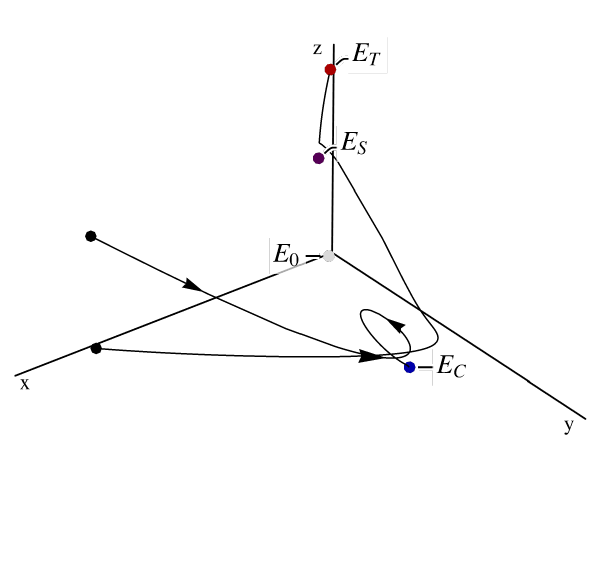} 
    \caption{\textbf{a} The curves $u=u_{TC}$ and $u=u_{SN}$ divide the parameter space $p\times u$ in regions $R_0$, $R_1$ and $R_2$ where system \eqref{sys2} have different equilibria configuration. These curves were plotted using their parametric expressions given in Section \ref{sec:proofs}. For each region, a representative parameter pair (black dots) is chosen and the corresponding phase portrait is plotted. The trajectories represent numerical solutions starting from initial conditions marked by black dots. Parameter values are given in Table \ref{tab:params}. \textbf{b} In region $R_0$, the tumor equilibrium $E_T$ is a stable node, while the nontrivial equilibria do not exist. Tumor escape is the only possible long-term outcome. \textbf{c} In region $R_1$, the tumor equilibrium $E_T$ is a saddle-point, while the tumor control equilibrium $E_C$ is the only positive nontrivial equilibrium, and it may be stable, as shown here. Tumor control is the only possible long-term outcome. \textbf{d} In region $R_2$, the tumor equilibrium $E_T$ is a stable node, the two nontrivial equilibria $E_C$ and $E_S$ are positive; in the case showed here, $E_C$ is stable and $E_S$ is a saddle with a two-dimensional stable manifold which is separatrix between the basins of attraction of $E_T$ and $E_C$. The long-term outcome (tumor escape or control) depends on where the initial condition is located.}
    \label{fig:regsefs}
\end{figure}

Moreover, considering $p$ (expansion rate of CAR-T cells) and $u$ (tumor immunosuppressive effect) as bifurcation parameters, in Theorem \ref{thmstb} we identify three regions in the parameter space where the equilibria present different existence and stability properties. These regions are separated by two curves, denoted as $u=u_{TC}(p)$ and $u= u_{SN}(p)$ (see Figure \ref{fig:regsefs}, expressions provided in Section \ref{sec:proofs}). The parameter regions are 
\begin{equation}
\begin{array}{l}
    {R}_0=\{
    \sigma\in\Lambda\;|\;
    p<p_*,\;u>u_{TC}
    \}\cup \{
    \sigma\in\Lambda\;|\;
    p>p_*,\;u>u_{SN}
    \},
    \\
    {R}_1=\{
    \sigma\in\Lambda\;|\;
    u<u_{TC}
    \}, \\
    {R}_2=\{
    \sigma\in\Lambda\;|\;
    p>p_*,\;u_{TC}<u<u_{SN}
    \}.
    \label{eq:regsR}
\end{array} 
\end{equation}

Within this context, the results of Theorem \ref{thmstb} are summarized as follows:
\begin{itemize}
    \item If $\sigma\in R_0$, then the tumor escape equilibrium $E_T$ is a stable node and there are no positive nontrivial equilibria.
    \item If $\sigma\in R_1$, then the tumor escape equilibrium $E_T$ is a saddle-point, the tumor control equilibrium $E_C$ is the only positive nontrivial equilibrium and it may be a stable or unstable.
    \item If $\sigma\in R_2$ then the tumor escape equilibrium $E_T$ is a stable node, both nontrival equilibria $E_C$ and $E_C$ are positive, $E_S$ is not stable while the tumor control equilibrium $E_C$ may be a stable or unstable.
\end{itemize}

The division of the parameter plane $p\times u$ into regions $R_0$, $R_1$ and $R_2$ is shown in Figure \ref{fig:regsefs}. One can interpret $R_0$ as a region of high immunosuppression and limited CAR-T cell expansion, where tumor escape is inevitable ($E_T$ is the only attractor); $R_1$ as a region of high CAR-T cell expansion and limited immunosuppression, where the tumor is likely to be under control ($E_C$ may be the only attractor), and $R_2$ as a region of relative balance between CAR-T cell expansion and immunosuppression, where the outcome (tumor escape or control) depends on the initial condition ($E_T$ and $E_C$ may be the only attractors). While the union $R_0 \cup R_1 \cup R_2$ does not cover the entire parameter space, it is \textit{dense} in $\Lambda$. The remaining parts correspond to the curves $u=u_{TC}$, where the tumor escape equilibrium $E_T$ undergo a transcritical with one of the nontrivial equilibria, and $u=u_{SN}$ where the nontrivial equilibria undergo a saddle-node bifurcation (see Section \ref{sec:proofs}; the intersection $u_{TC}=u_{SN}$ is a fold point).

Figure \ref{fig:regsefs} also shows representative phase portraits for parameter pairs $(p,u)$ in $R_0$, $R_1$ and $R_2$, in the particular cases where $E_C$ is stable in $R_1$ and $R_2$ and $E_S$ is a saddle in $R_2$, having a two-dimensional stable manifold. These stability properties hold for some parameter values, but not always. Thus, while the existence of the nontrivial equilibria is completely determined by the location of the parameter pair $(p,u)$, their stability cannot be completely determined in regions $R_1$ and $R_2$, as we show below. In Lemma \ref{lemmaB} we give sufficient conditions involving other parameters under which these properties seen in Figure \ref{fig:regsefs} hold.

\begin{table}[ht]
    \centering
    \begin{tabular}{|c|c|c|c|c|c|}
    \hline
    Figure & Region & $p$ & $u$ & $(x_0,y_0,z_0)$ & $(x_0,y_0,z_0)$ \\ \hline
    \ref{fig:regsefs}b, \ref{fig:plotts1}a & $R_0$ & $0.25$ & $0.2$ & $(0.3,0.001,0.6)$ & \\
    \ref{fig:regsefs}c, \ref{fig:plotts1}b & $R_1^s$ & $0.43$ & $0.2$ & $(0.05,0.001,0.6)$ & \\
    \ref{fig:regsefs}d, \ref{fig:plotts1}c,d & $R_2^s$ & $0.43$ & $0.3$ & $(0.05,0.001,0.6)$ & $(0.05,0.001,0.01)$ \\
    \ref{fig:regsefs2}b, \ref{fig:plotts2}a & $R_2^s$ & $0.66$ & $0.7$ & $(0.02,0.001,0.35)$ & $(0.03,0.03,0.3)$ \\
    \ref{fig:regsefs2}c, \ref{fig:plotts2}b & $R_2^u$ & $0.76$ & $0.7$ & $(0.02,0.001,0.35)$ & $(0.03,0.001,0.3)$ \\
    \ref{fig:regsefs2}d & $R_1^u$ & $0.76$ & $0.4$ & $(0.02,0.001,0.35)$ &  \\   
    \ref{fig:regsefs3}b & $R_2^{u,o}$ & $1.6568$ & $2.36$ & $(0.02,0.001,0.35)$ & $(0.03,0.001,0.3)$ \\
    \ref{fig:regsefs3}c, \ref{fig:plotts2}c & $\cal{L}$ & $1.6568$ & $2.7991$ & $(0.02,0.001,0.35)$ & $(0.03,0.001,0.3)$ \\
    \ref{fig:regsefs3}d, \ref{fig:plotts2}d & $R_2^{u,e}$ & $1.6568$ & $3.24$ & $(0.02,0.001,0.35)$ & $(0.019,0.06,0.16)$ \\ \hline  
    \end{tabular}
    \caption{The different values for $p$, $u$ and $(x_0,y_0,z_0)$, used in Figures \ref{fig:regsefs}-\ref{fig:plotts2}, are indicated here, while the other parameters were always set to $v=289/20,\ w=1/2,\ m=k=c=3/20,\ q=s=9/20$.}
    \label{tab:params}
\end{table}

\begin{figure}
    \begin{center}
    \begin{minipage}{0.8\linewidth}
    \textbf{\large{a}}\vspace{-0.1cm}\\
    \includegraphics[width=\linewidth]{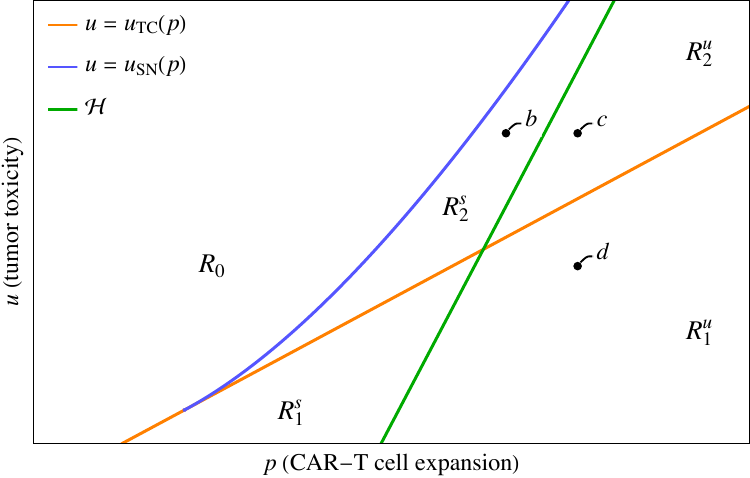}
    \end{minipage}
    \end{center}
    \vspace{0.4cm}
    
    \textbf{\large{b}} \hspace{4.5cm} {\textbf{\large{c}}} \hspace{4.5cm} {\textbf{\large{d}}} \vspace{-0.3cm}\\
    \includegraphics[width=0.32\linewidth]{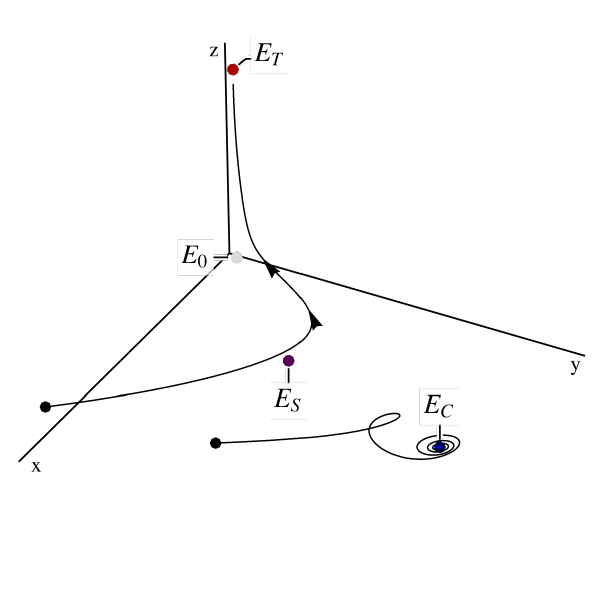}
    \includegraphics[width=0.32\linewidth]{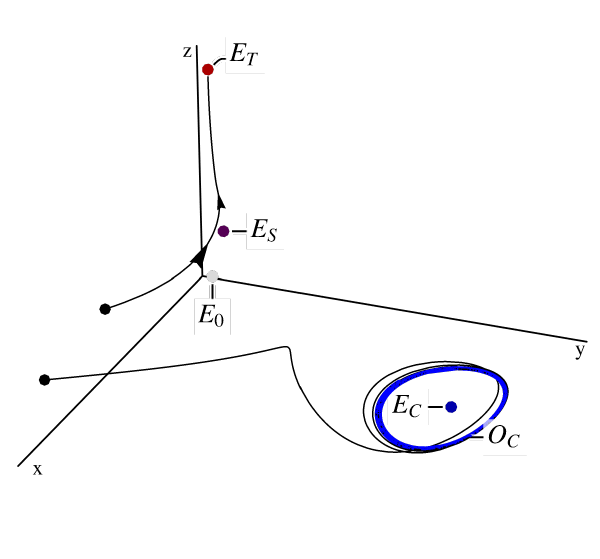}
    \includegraphics[width=0.32\linewidth]{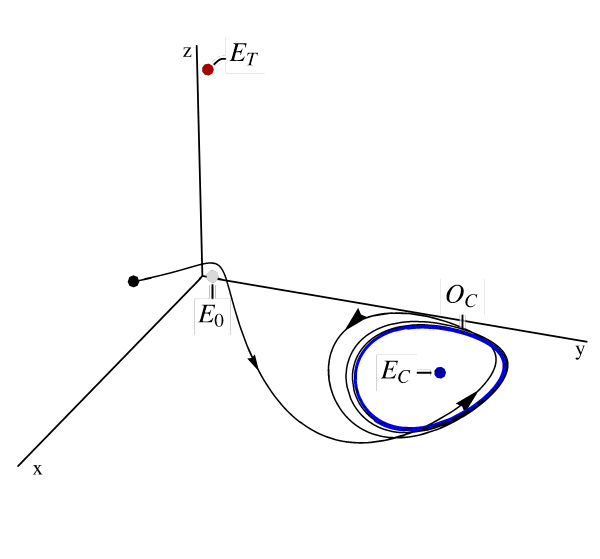} 
    \caption{\textbf{a} The curve $\cal{H}$ divides regions $R_1$ and $R_2$ in regions $R_i^s$ (where $E_C$ is stable) and $R_i^u$ (where $E_C$ is unstable, encircled by a small limit cycle for parameters close to $\cal{H}$). This curve was plotted using the parametric expression given in Section \ref{sec:proofs}. For each region, a representative parameter pair (black dots) is chosen and the corresponding phase portrait is plotted. The trajectories represent numerical solutions starting from initial conditions marked by black dots. Parameter values are given in Table \ref{tab:params}. \textbf{b} In region $R_2^s$, the tumor equilibrium $E_T$ is a stable node, $E_S$ is a saddle-point with a two-dimensional stable manifold, and the tumor control equilibrium $E_C$ is a stable focus in the case showed here. \textbf{c} Increasing $p$ leads to a Hopf bifurcation when the parameter pair $(p,u)$ crosses $\cal{H}$ from $R_2^s$ to $R_2^u$, where $E_C$ is now an unstable focus encircled by a stable limit cycle $O_C$ (blue trajectory). Tumor control is now achieved with sustained oscillation around a low tumor burden. \textbf{d} Decreasing $u$ leads to a transcritical bifurcation between $E_S$ and $E_T$ when $u$ crosses $u=u_{TC}(p)$ from $R_2^u$ to $R_1^u$, where now $E_T$ is a saddle and $E_C$ is still an unstable focus encircled by a stable limit cycle $O_C$ (blue trajectory).}
    \label{fig:regsefs2}
\end{figure}

In particular, the statement that the tumor control equilibrium $E_C$ may be stable or unstable suggests that it undergoes a Hopf bifurcation. Investigating this issue we rigorously proved in Theorem \ref{thmhop} that there exists a Hopf curve $\cal{H}$ characterized by the Jacobian of \eqref{sys2} evaluated at $E_C$ having a pair of complex eigenvalues with zero real part. We showed that the Hopf curve $\cal{H}$ divides the regions $R_1$ and $R_2$ into two subregions, denoted as $R_i^s$ and $R_i^u$ ($i=1,2$), where the superscripts $s$ and $u$ stand for stable and unstable, respectively (see Figure \ref{fig:regsefs2}). We prove that $E_C$ is stable for $\sigma \in R_i^s$ and that when $\sigma$ crosses $\cal{H}$ from $R_i^s$ to $R_i^u$, $E_C$ undergoes a Hopf bifurcation and becomes an unstable focus. Moreover, there exists a small limit cycle around $E_C$, denoted by $O_C$, for parameter values close to the curve $\cal{H}$, in one side $R_i^s$ or $R_i^u$. In all of our numerical results, the Hopf bifurcation was always supercriticial, meaning that $O_C$ is stable and arises in $R_i^u$; these results were later confirmed by Theorem \ref{thmbt} (see Section \ref{sec:proofs}). Figure \ref{fig:regsefs2} presents representative phase portraits of each case. Roughly speaking, we see that $\cal{H}$ defines, for each $u$ fixed, a threshold $p_H$ for the CAR-T cell expansion rate $p$, with respect the stability of $E_C$: if $p<p_H$ then $\sigma \in R_i^s$ and $E_C$ is stable, whereas if $p>p_H$ then $\sigma \in R_i^u$ and $E_C$ is unstable and encircled by a limit cycle, leading to sustained oscillations of small amplitude around the tumor control equilibrium.

\begin{figure}
    \begin{center}
    \begin{minipage}{0.8\linewidth}
    \textbf{\large{a}}\vspace{-0.1cm}\\
    \includegraphics[width=\linewidth]{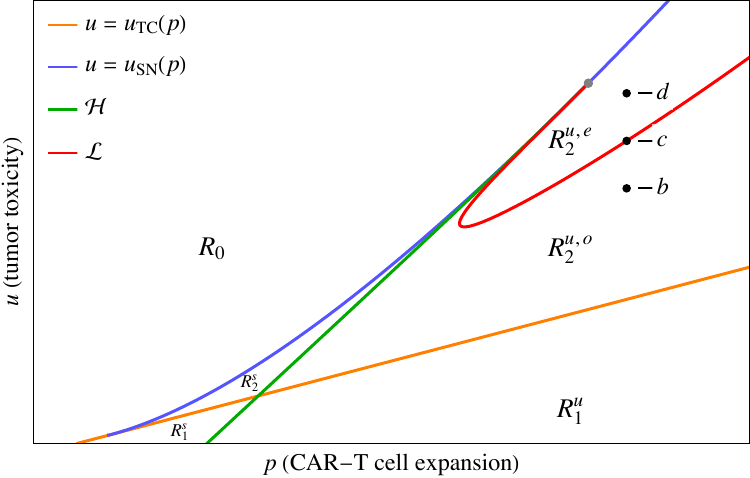}
    \end{minipage}
    \end{center}
    \vspace{0.4cm}
    
    \textbf{\large{b}} \hspace{4.5cm} {\textbf{\large{c}}} \hspace{4.5cm} {\textbf{\large{d}}} \vspace{-0.3cm}\\
    \includegraphics[width=0.32\linewidth]{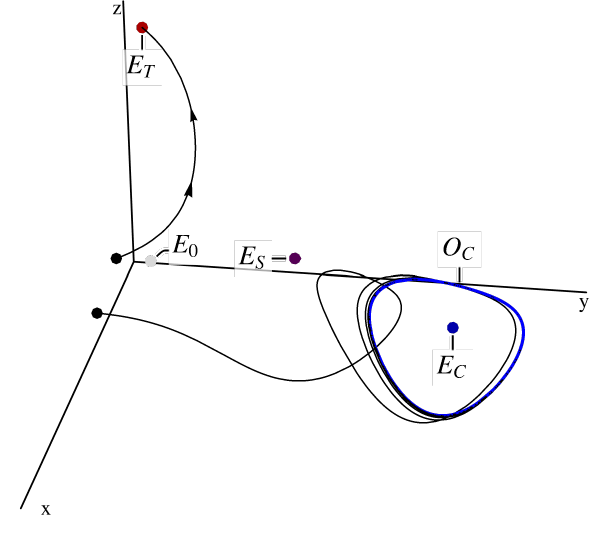}
    \includegraphics[width=0.32\linewidth]{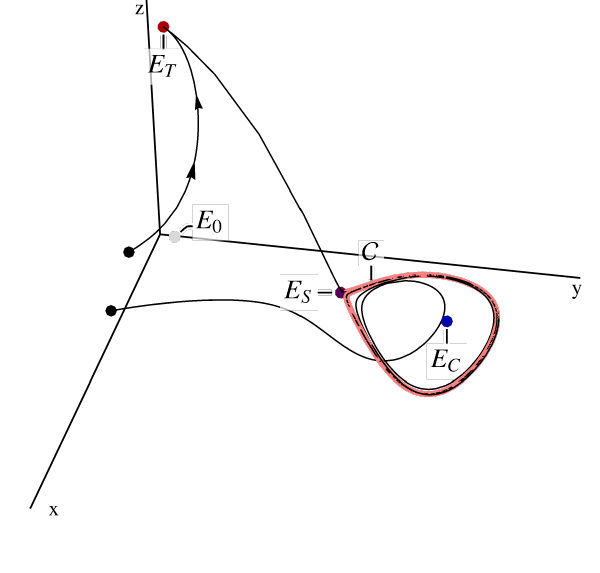}
    \includegraphics[width=0.32\linewidth]{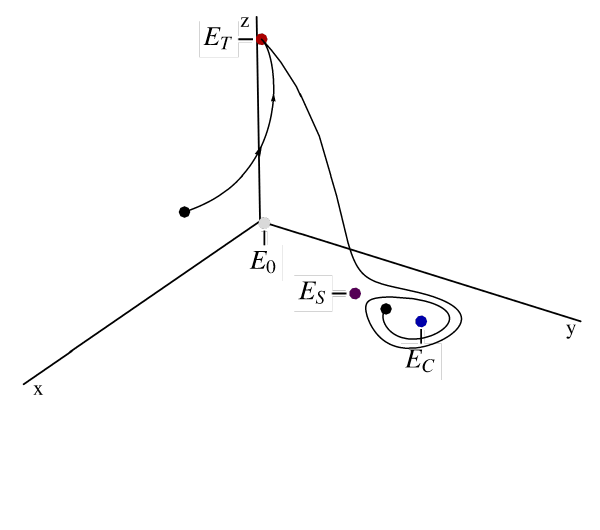} 

    \caption{\textbf{a} The curve $\cal{L}$ divides the region $R_2^u$ into the regions $R_2^{u,o}$ and $R_2^{u,e}$, where $o$ and $e$ stand for oscillation and escape, respectively. This curve was obtained numerically using MATCONT. For each region, a representative parameter pair (black dots) is chosen and the corresponding phase portrait is plotted. The trajectories represent numerical solutions starting from initial conditions marked by black dots. Parameter values are given in Table \ref{tab:params}. \textbf{b} In region $R_2^{u,o}$, the tumor equilibrium $E_T$ is a stable node, $E_S$ is a saddle with a two-dimensional stable manifold, and the tumor control equilibrium $E_C$ is an unstable focus encircled by a stable limit cycle $O_C$ (blue curve). The long term results are either tumor escape or tumor control with small amplitude oscillations. \textbf{c} Increasing the tumor toxicity $u$ leads to a global bifurcation where the limit cycle $O_C$ collapses with the saddle point $E_S$, forming a homoclinic loop $\cal{C}$ (pink curve). Solutions that appear to converge to the homoclinic loop escape the tumor control region and converge to the tumor escape equilibrium point $E_T$. Tumor escape is the only possible long-term outcome, which may be delayed when the solutions pass through the region of the unstable tumor control equilibrium. \textbf{d} Increasing $u$ leads to the disappearance of the homoclinic loop $\cal{C}$. Solutions passing near the tumor control equilibrium point $E_C$ escape with fewer oscillations around it and converge to the tumor escape equilibrium point $E_T$.}
    \label{fig:regsefs3}
\end{figure}

In Figure \ref{fig:regsefs2} we note that the Hopf curve $\cal{H}$ may intersect the saddle-node curve $u=u_{SN}$. At the intersection point, the Jacobian matrix of \eqref{sys2} evaluated at the tumor control equilibrium point $E_C$ has a pair of purely imaginary eigenvalues that collapse to a double zero eigenvalue. This indicates the presence of a Bogdanov-Takens bifurcation. We have investigated this aspect from a rigorous mathematical point of view and proved in Theorem \ref{thmbt} that indeed such an intersection exists and is a Bogdanov-Takens point where $E_C$ undergoes such a bifurcation. We give a full analytical proof, which was possible by imposing simplifying conditions on the model parameters. Figure \ref{fig:regsefs3} shows the corresponding two-dimensional bifurcation diagram.

The occurrence of a Bogdanov-Takens bifurcation implies the existence of a curve $\cal{L}$ of homoclinic loops in the parameter space (red curve in Figure \ref{fig:regsefs3}), i.e., a curve along which there is a homoclinic connection at the saddle-point $E_S$. We see that $\cal{L}$ divides the region $R_1^u$ into two subregions, denoted $R_1^{u,o}$ and $R_1^{u,e}$, where the superscripts $o$ and $e$ stand for oscillation and escape, respectively. In the region $R_1^{u,o}$ we have the same behavior as previously described for $R_1^{u}$: the tumor escape equilibrium is stable, the tumor control equilibrium is an unstable focus surrounded by a stable limit cycle $O_C$, and $E_S$ is a saddle with a two-dimensional stable manifold separating the basins of attraction of $E_T$ and $O_C$. As the parameter pair $(p,u)$ approaches the bifurcation curve $\cal{L}$, the limit cycle approaches the saddle point $E_S$, leading to a global bifurcation at $(p,u)\in \cal{L}$, when a homoclinic loop $\cal{C}$ is formed. When passing to $R_2^{u,e}$, the homoclinic loop disappears and there is no more attractor (either a cycle or an equilibrium) corresponding to tumor control. Solutions passing or starting near the tumor control equilibrium point $E_C$ spiral along it but then escape such region, converging to the tumor escape equilibrium $E_T$. This means that, although there is a tumor control equilibrium $E_C$ in region $R_2^{u,e}$, it cannot prevent tumor escape, but only delays it when the solutions travel through the region of $E_C$.

\subsection{Therapeutic implications}

\begin{figure}
    \begin{center}
    \includegraphics[width=0.2\linewidth]{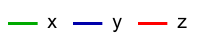}    
    \end{center}
    \vspace{-0.25cm}
    
    \textbf{\large{a}} \hspace{7.5cm} {\textbf{\large{b}}} \vspace{0.05cm}\\
    \includegraphics[width=0.48\linewidth]{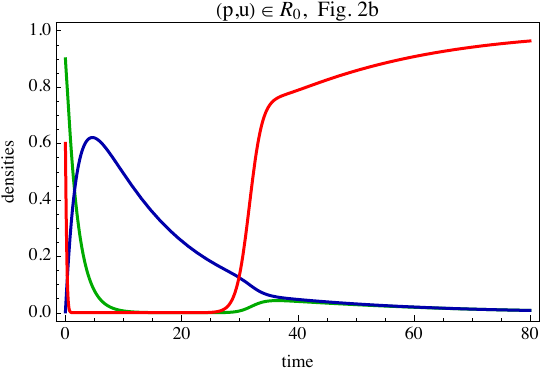}
    \hfill
    \includegraphics[width=0.48\linewidth]{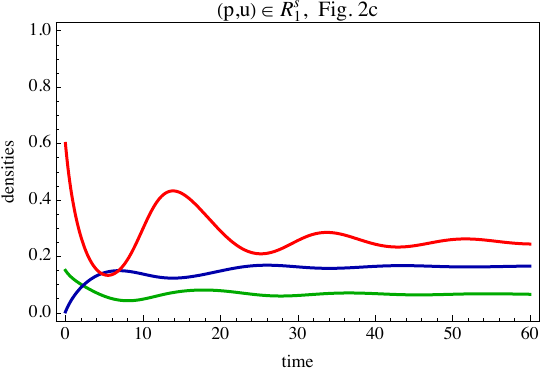}
    
    \textbf{\large{c}} \hspace{7.5cm} {\textbf{\large{d}}} \vspace{0.05cm}\\
    \includegraphics[width=0.48\linewidth]{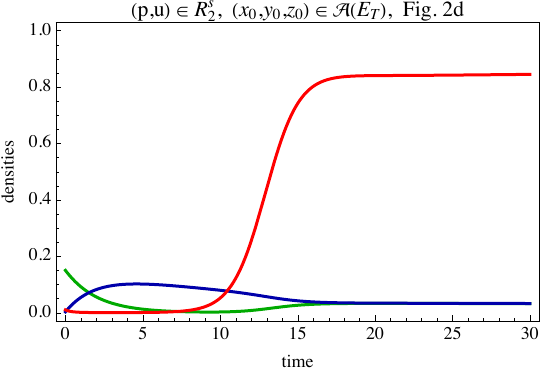}
    \hfill
    \includegraphics[width=0.48\linewidth]{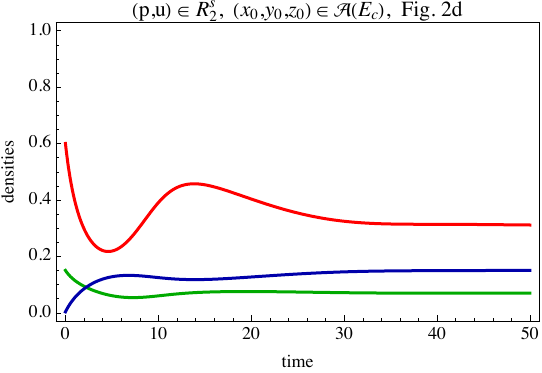}
    
    \caption{Time-courses of numerical solutions shown in Figure \ref{fig:regsefs} representing the different dynamics in the different parameter regions. Effector CAR-T cells are shown in green ($x$), memory CAR-T cells are shown in blue ($y$), and tumor cells are shown in red ($z$). Parameter values are given in Table \ref{tab:params}. \textbf{a} With high immunosuppression and low CAR-T cell expansion ($(p,u) \in R_0$), tumor escape is expected for all initial conditions, even after an initial good response. \textbf{b} For limited immunosuppression and CAR-T cell expansion ($(p,u) \in R_1^s$) tumor control with an equilibrium is expected for all initial conditions. \textbf{c,d} For intermediate immunosuppression and CAR-T cell expansion ($(p,u) \in R_2^s$), either tumor escape (c) or control with equilibrium (d) can occur, depending on which of the basins of attraction (denoted as $\cal{A}$$(E)$) the initial conditions are. Importantly, higher values of initial tumor burden seem to better stimulate the CAR-T cell expansion as required for tumor control.}
    \label{fig:plotts1}
\end{figure}

\begin{figure}
    \begin{center}
    \includegraphics[width=0.2\linewidth]{fig_plot_leg.pdf}    
    \end{center}
    \vspace{-0.25cm}
    
    \textbf{\large{a}} \hspace{7.5cm} {\textbf{\large{b}}} \vspace{0.05cm}\\
    \includegraphics[width=0.48\linewidth]{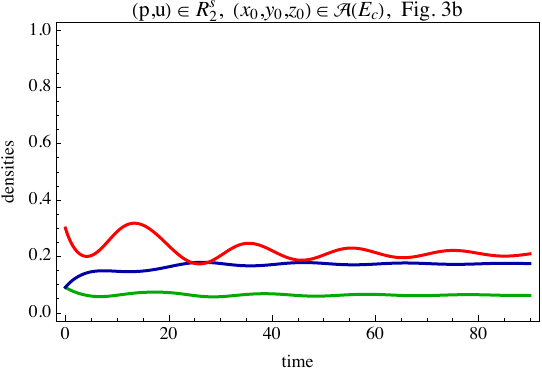}
    \hfill
    \includegraphics[width=0.48\linewidth]{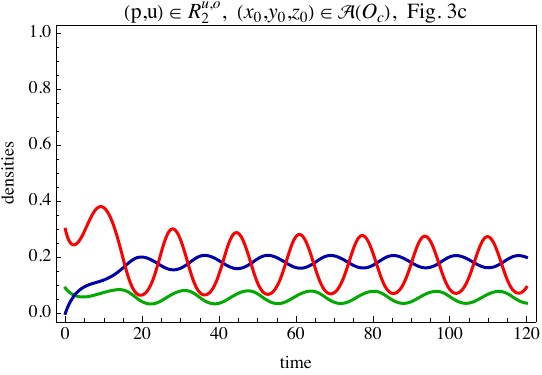}
    
    \textbf{\large{c}} \hspace{7.5cm} {\textbf{\large{d}}} \vspace{0.05cm}\\
    \includegraphics[width=0.48\linewidth]{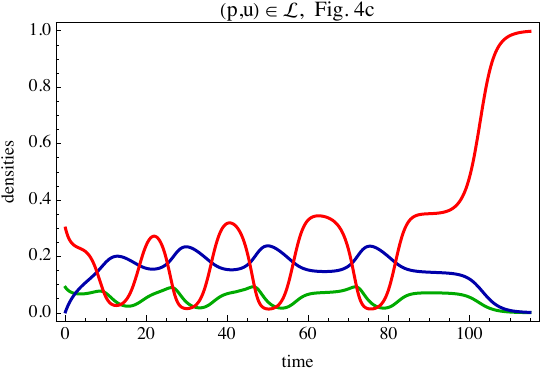}
    \hfill
    \includegraphics[width=0.48\linewidth]{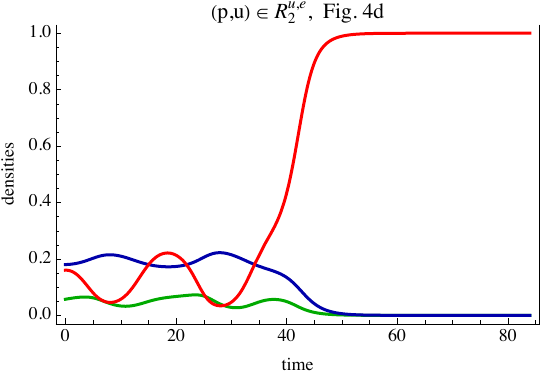}
    
    \caption{Time-courses of numerical solutions shown in Figures \ref{fig:regsefs2} and \ref{fig:regsefs3} representing the different dynamics in the different parameter regions. Effector CAR-T cells are shown in green ($x$), memory CAR-T cells are shown in blue ($y$), and tumor cells are shown in red ($z$). Parameter values are given in Table \ref{tab:params}. \textbf{a,b} The stable tumor control equilibrium (\textbf{a}) can be desestabilized by increased CAR-T cell expansion, leading to tumor control with sustained oscillations (\textbf{b}). \textbf{c,d} An even more desestabilizing effect is achieved by increasing immunosuppression, leading to a homoclinic bifurcation (c) where amplifying oscillations lead to tumor escape; this effect is stronger at higher immunosuppression (d).}
    \label{fig:plotts2}
\end{figure}

The above results have important implications for the therapeutic response to CAR-T cell treatment. We illustrate these consequences by discussing the time courses shown in Figures \ref{fig:plotts1} and \ref{fig:plotts2}, which correspond to the different trajectories shown in Figures \ref{fig:regsefs}, \ref{fig:regsefs2} and \ref{fig:regsefs3}.

Region $R_0$ is characterized by having the tumor escape as the only attractor, and represents a region of treatment failure due to the fact that tumor cells exert a strong immunosuppressive effect on CAR-T cells ($u>\max \{u_{TC},u_{SN}\}$). As shown in Figure \ref{fig:plotts1}a, a high dose of CAR-T cells can eventually reduce the tumor burden to very low levels, but this remission is only transient before the tumor escapes.

On the contrary, the region $R_1^s$ represents the ideal setting, since the tumor control equilibrium $E_C$ is the only attractor. This region requires a limited immunosuppressive effect of tumor cells ($u<u_{TC}$), but also a controlled expansion of CAR-T cells ($p<p_H$). In this case, even a small initial number of CAR-T cells is able to expand and suppress tumor growth, keeping a small but positive number of tumor cells under control, as shown in Figure \ref{fig:plotts1}b.

Region $R_2^s$ is characterized by bistability between tumor escape ($E_T$) and control ($E_C$). This configuration results from the relative limitation on the immunosuppressive effect of tumor cells and the expansion of CAR-T cells ($u_{TC}<u<u_{SN}$, $p<p_{H}$). In this bistable regime, it is important to note that the configuration of the basins of attraction may play an important and nontrivial role in the therapeutic outcome. Indeed, it seems that not all initial conditions consisting of small tumor loads lead to tumor control, even with higher CAR-T cell doses, but on the contrary, high tumor loads seem to be required. These claims are illustrated by the two solutions shown in the phase portrait in Figure \ref{fig:plotts1}d and their corresponding time courses in Figure \ref{fig:regsefs}c,d: while the initial condition $(x_0,y_0,z_0)=(0. 05,0.001,0.01)$ leads to tumor escape (Figure \ref{fig:regsefs}c), the same initial condition with $z_0=0.6$ leads to tumor control (Figure \ref{fig:regsefs}d). This complex configuration of the basins of attraction suggests that a high tumor burden may be required to activate CAR-T cells in order to achieve sufficient expansion that later leads to tumor control.

The occurrence of Hopf bifurcation within regions $R_1$ and $R_2$ also have important consequences for therapeutic implications, and is intrinsically related with the expansion rate of CAR-T cells. While tumor control consisting of exponential decay (Figure \ref{fig:plotts1}d) or damped oscillations (Figure \ref{fig:plotts2}a) is achieved when the CAR-T cell expansion is limited ($p<p_H$, $\sigma \in R_i^s$, $E_C$ is a stable node or focus), a high expansion rate ($p>p_H$, $\sigma \in R_i^u$), leads to an sustained oscillations of the residual tumor cells (the attractor is the limit cycle $O_C$ around $E_C$, Figure \ref{fig:plotts2}b). The results indicate that the high expansion rate of CAR-T cells leads to a stronger response against tumor cells, which then decline to very low levels. However, the stimulation of CAR-T cell expansion is then decreased, allowing tumor regrowth until the cycle repeats.

Finaly, the ocurrence of Bogdanov-Takens bifurcation suggests that a increased immunosuppressive effect $u$ of tumor cells may desestabilize tumor control by CAR-T cells, leading to transient control which only delays tumor escape. Indeed, we see in Figure \ref{fig:plotts2}c that, at the bifurcation curve $\cal{L}$, the collapse of the limit cycle $O_C$ in a homoclinic loop leads to amplifying oscillations of tumor cells until they reach a level where they are able to escape and suppress CAR-T cells. As the immunossuppressive effect increases (region $R_2^{u,e}$), this is effect becomes stronger, with less oscilations before escape (Figure \ref{fig:plotts2}d).

In summary, our analysis revealed the complex nonlinear dynamics of CAR-T cell therapy as described by the permanent system \eqref{sys2} and translated into important insights into long-term therapeutic outcome. In particular, we mention the nontrivial dependence on the initial tumor load required to achieve tumor control in the bistable regime; the role of increased CAR-T cell expansion in driving oscillating instead of damped tumor control, through a Hopf bifurcation; and the role of the immunosuppressive effect of tumor cells in transforming a scenario of oscillating tumor control into transient control followed by escape, through a Bogdanov-Takens bifurcation.

\section{Mathematical Analysis}
\label{sec:proofs}

In this section, we provide the formal statements, proofs and details for all results presented before.

\subsection{Stability of equilibria}

The following theorem summarizes the the existence and stability analysis of equilibria.

\begin{theorem}
\label{thmstb}
Consider system \eqref{sys2} with the parameter vector denoted by $\sigma=(p,u,\hat{\sigma})\in\Lambda$, $\hat{\sigma}=(v,w,m,k,q,s,c)$ fixed and consider $p$ and $u$ as bifurcation parameters. There exists $p_*>0$ and continuous increasing functions (denoted as bifurcation curves)
\begin{equation}
p\mapsto u=u_{TC}(p), \ p>0
\ \ \
\text{and}
\ \ \
p\mapsto u=u_{SN}(p),\ p\geqslant p_*
\end{equation}
satisfying $u_{TC}(p_*)=u_{SN}(p_*)$, $u_{TC}'(p_*)=u_{SN}'(p_*)$ and $u_{SN}(p)>u_{TC}(p)$ for $p\geqslant p_*$, and such that the following positivity and stability conditions for the equilibria of system \eqref{sys2} hold:
\begin{enumerate}
    \item The trivial equilibrium $E_0=(0,0,0)$ is a saddle; its two-dimensional stable manifold is the plane $z=0$, and its one-dimensional unstable manifold is the segment $0<z<1$ along the $z$-axis.
    \item The tumor escape equilibrium $E_T=(0,0,1)$ is always positive; it is a stable node if $u>u_{TC}$, and it is a saddle with a two-dimensional stable manifold if $u<u_{TC}$.
    \item There are up to two positive nontrivial equilibrium points, denoted by $E_C=(x_2,y_2,z_2)$ and $E_S=(x_3,y_3,z_3)$, which satisfy $0<z_2<z_3<1$ whenever both are positive. If $u<u_{TC}$ only $E_C$ is positive; if $p>p_*$ and $u_{TC}<u<u_{SN}$, $E_C$ and $E_S$ are positive; if $p>p_*$ and $u>u_{SN}$, or if $p<p_*$ and $u>u_{TC}$, there are no positive nontrivial equilibria. 
    \item If $J_S$ and $J_C$ denote the Jacobian matrix of system \eqref{sys2} evaluated at $E_S$ and $E_C$ respectively, then $J_S$ has at least one positive eigenvalue when $E_S$ is positive, and $J_C$ has at least one real negative eigenvalue when $E_C$ is positive.
\end{enumerate}

\end{theorem}

The proof of Theorem \ref{thmstb} is established below, where each paragraph concerns to one statement. The existence of $u_{TC}(p)$, $p_*$ and $u_{SN}(p)$ is established along the proof, and their expressions are given in \eqref{utc}, \eqref{ps} and \eqref{usn} respectively.

\paragraph{Invariance of  $\mathbb{R}^3_+$}
As a remark, it is straightforward to note that the first octant $\mathbb{R}^3_+$ is invariant with respect the flow of \eqref{sys2}. Therefore all trajectories with non-negative initial conditions remain with non-negative values.

\paragraph{Equilibrium points}
The equilibria of system \eqref{sys2} are easily obtained by setting the derivatives in \eqref{sys2} to zero. They are given by the trivial equilibrium $E_0=(0,0,0)$, the tumor escape equilibrium $E_T=(0,0,1)$, and up to two nontrivial equilibria, which are shown below.

\paragraph{Trivial equilibrium}
The stability of the trivial equilibrium $E_0$ is determined by the eigenvalues of the Jacobian matrix of \eqref{sys2} evaluated at $E_0$, which is
\[
J_0 = 
\left[
\begin{array}{ccc}
-w & 0 & 0 \\
s & -mq & 0 \\
0 & 0 & 1
\end{array}
\right]
.
\]
The eigenvalues are $-w$, $-mq$, and $1$. Therefore, $E_0$ is a saddle, with a two-dimensional stable manifold. This stable manifold is the invariant plane $z=0$, since $(0,0)$ is globally asymptotically stable for subsystem (\ref{sys2}a-b) with $z=0$ (this can be shown by solving $dx/d\tau =-wx$ and substituting the solution in $dy/d\tau = sx-mqy$). In a similar way, setting $x=y=0$ in \eqref{sys2}, we see that the unstable manifold of $E_0$ in $\mathbb{R}^3_+$ is the segment $0<z<1$ along the $z$-axis. This proves item 1 of Theorem \ref{thmstb}.

\paragraph{Tumor escape equilibrium}
The stability of the tumor escape equilibrium $E_T$ is determined by the eigenvalues of 
\[
J_T = 
\left[
\begin{array}{ccc}
\frac{p}{1+k}-u-w & q & 0 \\
s & -q-mq & 0 \\
-\frac{v}{1+c} & 0 & -1
\end{array}
\right]
.
\]
Denoting these eigenvalues by $\lambda_i^{(T)}$, $i=1,2,3$, we have $\lambda_3^{(T)}=-1<0$, while $\lambda_1^{(T)}$ and $\lambda_2^{(T)}$ are the eigenvalues of
\[
\hat{J}_T=
\left[
\begin{array}{cc}
\frac{p}{1+k}-u-w & q \\
s & -q(1+m)
\end{array}
\right]
.
\]
The characteristic polynomial of $\hat{J}_T$ is $\hat{P}(\lambda)=\lambda^2+b_1 \lambda +b_0$, with coefficients
\begin{equation}
b_0=q(1+m)\left(u-u_{TC} \right)
\ \ \
\text{and}
\ \ \
b_1=\left(u-u_{TC} \right) +(1+m)q+\frac{s}{m+1},
\label{coefsbi}
\end{equation}
where
\begin{equation}
    u_{TC}=\frac{p}{k+1}+\frac{s}{m+1}-w.
    \label{utc}
\end{equation}
Further, the discriminant of $\hat{P}(\lambda)$ is 
\begin{equation}
b_1^2-4b_0=
\left(u-u_{TC} - (1+m)q \right)^2+\frac{2s\left((u-u_{TC})+(1+m)q\right)}{m+1}+\frac{s^2}{(m+1)^2}.
\label{discbi}
\end{equation}
From \eqref{coefsbi}, \eqref{utc} and \eqref{discbi}, we conclude the following. If $u<u_{TC}$, then $b_0<0$ and $\lambda_1^{(T)}$ and $\lambda_2^{(T)}$ are real with opposite signs; since $\lambda_3^{(T)}<0$, in this case, $E_T$ is a saddle with a two-dimensional stable manifold. If $u>u_{TC}$, then $b_0>0$, $b_1>0$ and $b_1^2-4b_0>0$; thus $\lambda_1^{(T)}$ and $\lambda_2^{(T)}$ are real and negative, and $E_T$ is a stable node. This proves item 2 of Theorem \ref{thmstb}.

\paragraph{Nontrivial equilibria}

To find the nontrivial equilibria with $x,y,z>0$, we solve $dz/d\tau=0$ for $x$, obtaining
\begin{equation}
    \label{eqxc}
    x=x_C(z)=\dfrac{(1-z)(c+z)}{z+v-1}.
\end{equation}
Solving $dy/d\tau=0$ for $y$, we obtain
\begin{equation}
\label{eqyc}
y=y_C(z)=\dfrac{sx_C(z)}{q(z+m)}.
\end{equation}
Substituting $y=y_C(z)$ in $dx/d\tau =0$ with $x\neq 0$, we obtain
\begin{equation}
\phi(z)=0,
\ \ \
\text{where}
\ \ \
\phi(z)= \dfrac{p}{k+z}+\dfrac{s}{m+z}-\dfrac{w}{z}-u.
\label{equz}
\end{equation}
We conclude that the nontrivial equilibria have the form $(x_i,y_i,z_i)$ with
$$x_i=x_C(z_i) \ \ \text{and} \ \ y_i=y_C(z_i),$$
where $z_i$ is a real root of \eqref{equz}. A nontrivial equilibrium has biological meaning only when it is positive, i.e., if $x_i,y_i,z_i>0$. Since $v>1$, we have $x_i,y_i,z_i>0$ if and only if the root $z_i$ satisfies
\begin{equation}
0 <z_i<1,
\label{feascond}
\end{equation}
which we denote as feasibility conditions or feasibility interval.

Equation \eqref{equz} is equivalent to the third-degree polynomial equation 
\begin{equation}
\Gamma(z)=0, \ \ \ \text{where} \ \
\Gamma(z)= c_3z^3+c_2z^2+c_1z+c_0,
\label{polz}
\end{equation}
with coefficients
\begin{equation}
c_3= u,\ c_2=w+mu+ku-s-p, \ c_1= mku+kw+mw-ks-mp, \ c_0 = kmw.
\label{coefspol}
\end{equation}
Note that $c_3>0$ implies $\Gamma(-\infty)=-\infty$. Since $\Gamma(0)=c_0>0$, the Intermediary Value Theorem (IVT) guarantees that $\Gamma$ has a real negative root $z_1<0$. Thus, $\Gamma(z)$ admits at most two positive roots, which we denote by $z_2 $ and $ z_3$. When both roots are positive, we denote then in the order $z_2<z_3$. Therefore, there are up to two nontrivial equilibria, $E_C=(x_2,y_2,z_2)$ and $E_S=(x_3,y_3,z_3)$.

We analyze the existence of these roots assuming $u$ as bifurcation parameter. Note that
\begin{equation}
\Gamma(1)=(k+1)(m+1)\left(u-u_{TC}\right).
\end{equation}
Therefore, if $0<u<u_{TC}$, then $\Gamma(0)>0>\Gamma(1)$ and $\Gamma(1)<0<\Gamma(\infty)=\infty$. Thus, by the IVT, $\Gamma(z)$ has a root $z_2\in(0,1)$ and another root $z_3>1$. By \eqref{feascond}, we conclude that if $u\in (0,u_{TC})$ there is a unique positive nontrivial equilibrium, which is $E_C$.


Note that $\partial \Gamma/\partial u=kmz+(k+m)z^2+z^3>0$ for $z>0$. Therefore, as $u$ increases, the values of $\Gamma(z)$ also increase, for all $z>0$. Therefore, the roots $z_2<1$ and $z_3>1$ move to right and left respectively. When $u$ reaches $u_{TC}$, one of these roots $z_{\tilde{i}}$ reaches $z=1$ because $\Gamma(1)=0$ if $u=u_{TC}$. Note that this means a transcritical bifurcation between some of the $E_{{i}}$ ($i=C,S$) and $E_T$ at $u=u_{TC}$, since $z_{{i}}=1$ implies $E_{{i}}=(0,0,1)=E_T$. Note that, since $\Gamma$ is a third-degree polynomial with three real roots $z_1<0<z_2<z_3$ and leading coefficient $c_3>0$, we necessarily have $\Gamma'(z_1)>0$, $\Gamma'(z_2)<0$ and $\Gamma'(z_3)>0$. Therefore, to decide which root $z_2$ or $z_3$ first crosses $z=1$ as $u$ reaches $u_{TC}$, we analyze the sign of the derivative of $\Gamma(z)$ at $z=1$ when the crossing occurs, i.e., the sign of $\rho_*=\Gamma'(1)$ calculated with $u=u_{TC}$. If $\rho_*<0$, we have $z_2=1$ when $u=u_{TC}$; if $\rho_*>0$, we have $z_3=1$ when $u=u_{TC}$. The formula for $\rho_*$ is
\[
\rho_*=\frac{(m+1)}{k+1}\left( p- p_* \right),
\]
where
\begin{equation}
p_*=(k+1)^2 \left(w-\frac{s}{(m+1)^2}\right).
 \label{ps}
\end{equation}
Note that $p_*>0$, because $w>s$ (see \eqref{nondimpars}).

First, we assume $p<p_*$. If $u\in(0,u_{TC})$, we have a single root $z_2\in(0,1)$ and a single positive nontrivial equilibrium $E_C$; when $u=u_{TC}$, $E_C$ collides with $E_T$; for $u>u_{TC}$ the root $z_2$ leaves the feasibility interval \eqref{feascond} and there are no positive nontrivial equilibria.

Now, assuming $p>p_*$, we have the following. If $u\in (0,u_{TC})$, we have a root $z_2\in (0,1)$ and other root $z_3\in (1,\infty)$, thus there is a single positive nontrivial equilibrium $E_C$. For $u=u_{TC}$, we have $z_3=1$ and $E_S$ coincides with $E_T$. For $u$ slightly greater than $u_{TC}$, we have $0<z_3<1$ and the $E_S$ is now also positive. As $u$ increases, the two positive roots $z_2<z_3$ persist until colliding when $u$ reaches a critical threshold $u_{SN}>u_{TC}$, i.e., $E_C$ and $E_S$ collapse to a saddle-node point at $u_{SN}$ (see below). For $u>u_{SN}$, there is no positive nontrivial equilibria. This proves item 3 of Theorem \ref{thmstb}.

\paragraph{The bifurcation curves}
The threshold $u_{SN}$ and the collision point where $z_2=z_3$ are the solutions $(u,z)$ of the system of equations
\[
\left\lbrace
\begin{array}{l}
     \,\, \Gamma(z)=0,  \\
     \Gamma'(z)=0, \\
     0<z<1.
\end{array}
\right.
\]
Solving this system for $p$ and $u$, we obtain a parameterization $z\mapsto (p_{SN}(z),u_{SN}(z))$ in the plane $p\times u$ of a curve along which $\Gamma(z)$ has a double real root satisfying the feasibility conditions. The parametrized curve is given by
\begin{equation}
\left\lbrace
    \begin{array}{l}
    p=p_{SN}(z)=(k+z)^2 \left(\dfrac{w}{z^2}-\dfrac{s}{(m+z)^2}\right),\vspace{0.2cm} \\
    u=u_{SN}(z)=\dfrac{ (m-k)s}{(m+z)^2}+\dfrac{k w}{z^2},\vspace{0.2cm} \\
    0<z<1.
    \end{array}
    \label{paramupsn}
\right.
\end{equation}
We now prove some properties. First, note that $p_{SN}(z)>0$ since $w>s$ and
\begin{equation*}
    \dfrac{1}{z^2}-\dfrac{1}{(m+z)^2}>0.
    \label{desigzden2}
\end{equation*}
Further, from \eqref{paramupsn}, $u_{SN}(z)$ can also be written as
\begin{equation}
    u_{SN}(z)=k\dfrac{p_{SN}(z)}{(k+z)^2}+\dfrac{m s}{(m+z)^2}.
    \label{usnpz}
\end{equation}
Therefore, $u_{SN}(z)>0$. Further, we have
\[
\lim_{z\to 0^+} (p_{SN}(z),u_{SN}(z))=(+\infty,+\infty),
\ \ \text{and} \ \ 
\lim_{z\to 1^-} (p_{SN}(z),u_{SN}(z))=(p_*,u_{TC}(p_*)).
\]
Thus, the saddle-node curve belongs to the first quadrant of the plane $p\times u$, and connects a point at infinity to the point in the curve $u=u_{TC}(p)$ with $p=p_*$. Note also that
\begin{equation}
    p_{SN}'(z)=
-(k+z) \left(2 k s \left(\dfrac{1}{z^3}-\dfrac{1}{(m+z)^3}\right)+\dfrac{2 k (w-s)}{z^3}+\dfrac{2 m s}{(m+z)^3}\right)<0,
\label{derpsn}
\end{equation}
since $w>s$ and
\begin{equation*}
    \dfrac{1}{z^3}-\dfrac{1}{(m+z)^3}>0.
    \label{desigzden3}
\end{equation*}
Therefore, $p_{SN}(z)$ is a strictly decreasing function in the interval $z\in (0,1)$ whose image is $(p_*,\infty)$. By the Inverse Function Theorem, $p_{SN}(z)$ admits an inverse function $p \mapsto z_{SN}(p)$, defined for $p\in (p_*,\infty)$ and whose image is $(0,1)$, with derivative satisfying
\begin{equation}
    z_{SN}'(p)=\dfrac{1}{p_{SN}'(z)}<0.
\label{derzsnp}
\end{equation}
Substituting $z=z_{SN}(p)$ in \eqref{usnpz} gives the expression of $u_{SN}$ as function of $p$ only, 
\begin{equation}
    u_{SN}(p)=\dfrac{kp}{(k+z_{SN}(p))^2}+\dfrac{m s}{(m+z_{SN}(p))^2}.
    \label{usn}
\end{equation}
Differentiation shows that $u_{SN}(p)$ is a strictly increasing function, since $z_{SN}'(p)<0$ and
\begin{equation}
    u_{SN}'(p)=
    -z_{SN}'(p) \left(\frac{2 k p}{(k+z_{SN}(p))^3}+\frac{2 m s}{(m+z_{SN}(p))^3}\right)+\frac{k}{(k+z_{SN}(p))^2}>0.
    \label{derusnp}
\end{equation}
We note that $u_{SN}(p)$ and $u_{TC}(p)$ not only intersect at $p=p_*$ but are also tangent. Indeed, making $p\to p_*$ in \eqref{derusnp} we have $z_{SN}(p)\to 1$ and then using \eqref{derzsnp}, we have
\[
u_{SN}'(p_*)=
    -\dfrac{1}{p_{SN}'(1)}\left(\frac{2 k p_*}{(k+1)^3}+\frac{2 m s}{(m+1)^3}\right)+\frac{k}{(k+1)^2}.
\]
Substituting \eqref{derpsn} with $z\to 1$ in the above expression leads to the conclusion that
\[
u_{SN}'(p_*)=\dfrac{1}{k+1}=u_{TC}'(p_*).
\]
Finally, we show that $u_{SN}(p)>u_{TC}(p)$ for all $p>p_*$. Since $z\mapsto p_{SN}(z)$ in \eqref{paramupsn} is a one-to-one map from $(0,1]$ to $[p_*,\infty)$ (the extension to $z=1$ follows from the limits above), we see that writing $p=p_{SN}(z)$ in \eqref{utc} gives a parameterization of $u=u_{TC}(p)$ in terms of $z$, $z\mapsto (p_{SN}(z),u_{TC}(p_{SN}(z))) $. Since
\[
u_{SN}(z)-u_{TC}(p_{SN}(z))=
\dfrac{(z-1)^2 \left(k (m+1) m^2 w+z^2 (k (m w+w-s)+m s)+2 k (m+1) m w z\right)}{(k+1) (m+1) z^2 (m+z)^2},
\]
and $w>s$, we have $u_{SN}(z)>u_{TC}(p_{SN}(z))$ for all $z\in (0,1)$ and thus $u_{SN}(p)>u_{TC}(p)$ for all $p>p_*$.

The above results prove the statements in Theorem \ref{thmstb} about properties of the bifurcation curves $p\mapsto u_{TC}(p)$ and $p\mapsto u_{SN}(p)$.

\paragraph{Stability of nontrivial equilibria} 
We now analyze the stability of the nontrivial equilibrium $E_i$, $i=C,S$ under the assumption that they are positive, giving a proof for item 4 of Theorem \ref{thmstb}. We denote the Jacobian matrix of system \eqref{sys2} at $E_i$ as $J_i$, $i=C,S$. Using the fact that $z_i$ satisfies \eqref{equz} and $y_i$ satisfies \eqref{eqyc}, we simplify the entries $j_{11}$, $j_{13}$ and $j_{23}$, obtaining
\begin{equation}
J_i = 
\left[
\begin{array}{ccc}
-\dfrac{sz_i}{m+z_i} & qz_i & j_{13} \\
s & -q(m+z_i) & -\dfrac{sx_i}{m+z_i} \\
j_{31} & 0 & j_{33}
\end{array}
\right]
,
\label{jacobian}
\end{equation}
where
\begin{equation}
\label{jlks}
j_{13} = x_i \left( \frac{w}{z_i} - \frac{kpz_i}{(k+z_i)^2} \right),
\ \
j_{31} = -\frac{vz_i(c+z_i)}{(c+x_i+z_i)^2},
\ \
j_{33} = 1-2z_i-\frac{vx_i(c+x_i)}{(c+x_i+z_i)^2}
.
\end{equation}
The characteristical polynomial of $J_i$ is given by
\begin{equation}
P_E(\lambda)=\lambda^3+d_2\lambda^2 + d_1 \lambda+d_0,
\end{equation}
where
\begin{equation}
\begin{array}{l}
d_2=\dfrac{sz_i}{(m+z_i)}+q(m+z_i)-j_{33},
\vspace{0.2cm}\\
d_1=-j_{13} j_{31}-j_{33} \left(q (m+z_i)+\dfrac{ s z_i}{m+z_i}\right),
\vspace{0.2cm}\\
d_0=j_{31}q\left(\dfrac{sx_iz_i}{m+z_i}-j_{13}(m+z_i)\right).
\end{array}
\label{jacobiancoefs}
\end{equation}

We analyze the sign of $d_0$. Comparing the expression of $j_{13}$ with the derivative of $\phi(z)$ in \eqref{equz}, we can write
\[
j_{13} = x_i z_i \left( \frac{s}{(m+z_i)^2} +\phi'(z_i) \right).
\]
Substituting this and $j_{31}$ in $d_0$, we obtain
\begin{equation}
\label{eq:d0Jc}
d_0=\frac{qvx_iz_i^2(c+z_i) (m+z_i)}{(c+x_i+z_i)^2}\phi'(z_i).
\end{equation}
From the results on the existence of the nontrivial equilibria, we note that $z_2$ and $z_3$ are the unique positive roots of $\phi(z)=0$. Since $0<z_2<z_3$ and (see \eqref{equz})
\[
\lim_{z\to0^+}\phi(z)=-\infty
\ \ 
\text{and}
\ \ 
\lim_{z\to\infty}\phi(z)=-u,
\]
we conclude from the differentiability of $\phi$ that $\phi'(z_2)>0>\phi'(z_3)$ (otherwise, $\phi$ would have other roots). This implies that $d_0>0$ for $i=C$ ($z_i=z_2$) and $d_0<0$ for $i=S$ ($z_i=z_3$). Since $d_0=-\lambda_1
\lambda_2\lambda_3$, where $\lambda_j$ are the eigenvalues of $J_i$, we conclude that at least one eigenvalue of $J_C$ is real negative when $E_C$ is positive and at least one eigenvalue of $J_S$ is real positive when $E_S$ is positive. This concludes the proof of Theorem \ref{thmstb}.

\paragraph{Sufficient conditions for $d_2>0$}
While the above results ($d_0<0$ for $J_S$) implies that $E_S$ is not stable because $J_S$ has at least one positive eigenvalue, it does not establish that the remaining two eigenvalues have real negative parts, i.e., that $E_S$ is a saddle with a two-dimensional stable manifold. Note, however, that the condition $d_2>0$ implies this property, since $d_2=-(\lambda_1+\lambda_2+\lambda_3)$.

On the other hand, the above results also do not guarantee that $E_C$ is stable. To prove this, the Routh-Hurwitz conditions
\[
d_0>0, \ \
d_2>0 \ \ 
\text{and}
\ \ 
d_2d_1-d_0>0,
\]
must be checked for $J_C$. While $d_0>0$ is always satisfied, the remaining conditions require additional hypotheses on the parameters. This implies that $E_C$ \textit{may} be stable but not always, suggesting a Hopf bifurcation when $d_2d_1-d_0=0$.

In light of these remarks, we now give conditions which imply $d_2>0$ for $J_S$ and $J_C$. We will also use these conditions in Theorem \ref{thmhop} to show that $E_C$ undergoes a Hopf bifurcation.

To express $d_2$ in terms of $z_i$ and the model parameters, we replace $x_i=x_C(z_i)$ from \eqref{eqxc} in $j_{33}$ in \eqref{jlks} and then in $d_2$ in \eqref{jacobiancoefs}, obtaining
\begin{equation}
    d_2 =
    \frac{s z_i}{m+z_i}+q (m+z_i)+\frac{z_i (z_i-1)^2}{v (c+z_i)}+\frac{z_i (c+2 z_i-1)}{(c+z_i)}.
    \label{d2z}
\end{equation}
The first three terms in \eqref{d2z} are positive. Note that the roots $z_i$ depend only on $p,k,s,m,w$ and $u$ (see \eqref{equz}). Therefore, once these parameters are given and determine $z_i$, if $c>1-2z_i$ we have $d_2>0$. If this is not true, then we see that there are thresholds for $q$ and $v$ so that $d_2>0$. These conclusions are summarized in the following Lemma.

\begin{lemma} Given $p,k,s,m,w,u$, and $z_2$ and $z_3$ the two possible real and positive roots of \eqref{equz}, if one of the following conditions holds, then the coefficient $d_2$ of the characteristic polynomials of $J_S$ and $J_C$ is positive:
\begin{enumerate}
    \item $c>1-2z_i$,
    \item $q>\dfrac{1}{(m+z_i)} \left( 
    \dfrac{z_i (1-c-2 z_i)}{(c+z_i)}-\dfrac{s z_i}{m+z_i}-\dfrac{z_i (z_i-1)^2}{v (c+z_i)}
    \right)$,
    \item $\dfrac{1}{v}>\dfrac{(c+z_i)}{z_i (z_i-1)^2} \left( 
    \dfrac{z_i (1-c-2 z_i)}{(c+z_i)}-\dfrac{s z_i}{m+z_i}-q (m+z_i)
    \right)$,
    \item $q$ and $1/v$ are large enough so that $q (m+z_i)+\frac{1}{v}\frac{z_i (z_i-1)^2}{c+z_i}>
    \frac{z_i (1-c-2 z_i)}{(c+z_i)}-\frac{s z_i}{m+z_i}$.
\end{enumerate}
    \label{lemmaB}
\end{lemma}

In addition to the conditions of Lemma \ref{lemmaB}, if one also admits that the third Routh-Hurwitz condition holds for $J_C$, then we have the scenario shown in Figure \ref{fig:regsefs} and given in the following Corollary.

\begin{corollary} Under the conditions of Lemma \ref{lemmaB}, $E_S$ is a saddle with a two-dimensional stable manifold, while $E_C$ is stable if $d_2d_1-d_0>0$ also holds.
    \label{corollaryB}
\end{corollary}

\subsection{Hopf bifurcation}

As shown above, it is not always the case that $E_C$ is stable, and our previous results suggest that it can become unstable via a Hopf bifurcation. We now briefly recall some aspects of Hopf bifurcation theory and present a simple criterion by Liu that gives sufficient and necessary conditions for a \textit{simple} Hopf bifurcation. We then use this approach to prove that the tumor control equilibrium $E_C$ undergoes such a bifurcation. 

A Hopf bifurcation is a local bifurcation in which an equilibrium of a dynamical system loses stability because a pair of complex conjugate eigenvalues of the Jacobian matrix crosses the imaginary axis \cite{hassard1981theory,guckenheimer1986dynamical,perko2013differential,Kuznetsov}. More specifically, consider system
\begin{equation}\label{eq:ED2}
    \mathbf{x}'=f(\mathbf{x},\zeta),
\end{equation}
where $\mathbf{x}=(x,y,z) \in U\subset \mathbb{R}^3$, $f$ is of class $C^{\infty}$ on the open set $U$, and $\zeta\in \mathbb{R}$ is a bifurcation parameter, while the other parameters are fixed. Suppose it has an equilibrium $\mathbf{x}_0=\mathbf{x}_0(\zeta)$ that is a function of $\zeta$ for $\zeta$ near certain $\zeta_H$. Denote the Jacobian matrix of \eqref{eq:ED2} by $J(\mathbf{x},\zeta)$. If at $\zeta=\zeta_H$, $J(\mathbf{x}_0,\zeta_H)$ has a simple pair of purely imaginary eigenvalues $\pm \lambda\left(\zeta_H\right)$ and no other eigenvalues with zero real parts and if $\left.\frac{d}{d \zeta} \operatorname{Re} \lambda(\zeta)\right|_{\zeta=\zeta_H} \neq 0$, then a small-amplitude periodic solution surrounding $\mathbf{x}_0$ emerges as $\zeta<\zeta_H$ or $\zeta>\zeta_H$, for $\zeta$ is close to $\zeta_H$. The particular case where all other eigenvalues (one, in the third-dimensional case) of $J(\mathbf{x}_0,\zeta_H)$ have negative real parts is called \textit{simple Hopf bifurcation} \cite{liu1994criterion,chen2021collective,shih2022hopf}.
    
Typically, the criteria for a Hopf bifurcation of codimension 1 involve calculating the eigenvalues of the Jacobian and the normal form of the system around the equilibrium. The computation of the Lyapunov coefficient, which determines whether the bifurcation is subcritical, supercritical, or degenerate, is even more complex, requiring the computation of second and third order coefficients around equilibrium. This task is challenging for high-dimensional systems, and even for low-dimensional systems where the equilibria depend on roots of higher-degree polynomials. In such cases, theoretical analysis with symbolic computations becomes difficult, if not impossible.

An alternative and simpler approach to guarantee the existence of a Hopf bifurcation was proposed by Liu in 1994 \cite{liu1994criterion} and has been applied in various modeling contexts \cite{chen2021collective,shih2022hopf,liu1987dynamical,liu1997nonlinear,domijan2009bistability}. This criterion is based on the degeneracy of one of the Routh-Hurwitz conditions and relies solely on the coefficients of the characteristic polynomial of the Jacobian. Although it only guarantees the existence of a simple Hopf bifurcation and does not provide information about its nature (subcritical, supercritical, or degenerate), it is very useful for identifying the locus (manifold) in the parameter space where the equilibrium has a pair of purely imaginary eigenvalues.

For simplicity, we state Liu's theorem as recently presented by Shih \cite{shih2022hopf}, restricting ourselves to the dimension $n=3$. For a comprehensive description of this approach, refer to  \cite{chen2021collective}.

\begin{theorem}
    \label{thm:hopf}
    \cite{shih2022hopf} Consider system \eqref{eq:ED2} and denote the characteristic polynomial of $J({\mathbf{x}},\zeta)$ by
    \begin{equation}
        P_E(\mathbf{x},\zeta,\lambda)=\lambda^3+d_2(\mathbf{x},\zeta)\lambda^2 + d_1(\mathbf{x},\zeta) \lambda+d_0(\mathbf{x},\zeta).
        \label{eq:charac_polynomial_PE}
    \end{equation}
    A simple Hopf bifurcation occurs at $\mathbf{x}=\mathbf{x}_0=\mathbf{x}_0\left(\zeta_H\right)$ and $\zeta=\zeta_H$ if and only if
    \begin{equation}
    \begin{aligned}
    & \,d_0\left(\mathbf{x}_0,\zeta_H\right)>0,\\
    & \,d_2\left(\mathbf{x}_0,\zeta_H\right)>0,\\
    & \,d_2\left(\mathbf{x}_0,\zeta_H\right)d_1\left(\mathbf{x}_0,\zeta_H\right)-d_0\left(\mathbf{x}_0,\zeta_H\right)=0,\\
    & \left.\frac{{d}}{{d} \zeta}\left[d_2\left(\mathbf{x}_0,\zeta\right)d_1\left(\mathbf{x}_0,\zeta\right)-d_0\left(\mathbf{x}_0,\zeta\right)\right]\right|_{\zeta=\zeta_H} \neq 0 .
    \end{aligned}
    \label{conditionsLiu}
    \end{equation}
\end{theorem}

We apply Theorem \ref{thm:hopf} to system \eqref{sys2} and the tumor control equilibrium $E_C$. The Jacobian matrix evaluated at $\mathbf{x}_0=E_C=(x_3,y_3,z_3)$, and the coefficients of the characteristic polynomial $P_E(\lambda)$ are given in \eqref{jacobian} and \eqref{jacobiancoefs}. We proved that $d_0>0$, and, under the conditions of Lemma \ref{lemmaB}, we have $d_2>0$. According to Theorem \ref{thm:hopf}, the Hopf locus, i.e., the manifold in the parameter space where a Hopf bifurcation occurs, is given by
\[
\psi:=d_2d_1-d_0=0.
\]

To obtain a characterization of the Hopf locus in the parameter plane $p\times u$, we proceed in the same way as before for the saddle-node bifurcation, obtaining a parameterization $z\mapsto (p_H(z),u_H(z))$ of the curve $\psi=0$, keeping the other parameters fixed. This parameterization is the solution $(p,u)$ of the system of equations
\begin{equation}
\label{sishopf}
\left\lbrace
\begin{array}{l}
     \Gamma(p,u,z)=0,  \\
     \psi(p,u,z)=0.
\end{array}
\right.
\end{equation}
Indeed, for each $z$, the solution of this system correspond a parameter pair $(p_H(z),u_H(z))$ for which $z$ is a root of $\Gamma(z)=0$ (equation \eqref{polz}) and $\psi=0$. To solve \eqref{sishopf} for $p$ and $u$ in terms of $z$, we note the following. First, from \eqref{jacobiancoefs}, we have
\[
\psi
=
j_{13} j_{31} q (m+z)-\frac{j_{31} q s x z}{m+z}+\left(j_{33}-q (m+z)-\frac{s z}{m+z}\right) \left(j_{13} j_{31}+j_{33} \left(q (m+z)+\frac{s z}{m+z}\right)\right).
\]
Checking the expressions of $j_{13}$, $j_{31}$ and $j_{33}$ in \eqref{jlks}, we see that none of them depend on $u$, and only $j_{13}$ depends on $p$, being a degree one polynomial in $p$. Therefore, $\psi$ is also a degree one polynomial in $p$. Replacing $x_i$ by \eqref{eqxc} with $z=z_i$, we can write
\[
\psi
=
A_\psi(z) + B_\psi(z) p,
\]
where
\[
\begin{array}{l}
A_\psi(z)
=
\frac{j_{31} (1-z) (c+z) (j_{33} w (m+z)-s z (q z+w))}{z (m+z) (v+z-1)}+j_{33} \left(q (m+z)+\frac{s z}{m+z}\right) \left(j_{33}-q (m+z)-\frac{s z}{m+z}\right),
\vspace{0.2cm} \\
B_\psi(z)
=
\frac{j_{31} (1-z) z (c+z)}{(k+z)^2 (v+z-1)}\left(\frac{sz}{m+z}-j_{33}\right).
\end{array}
\]
With this, the second equation of \eqref{sishopf} is easily solvable for $p$, while the first is then solved for $u$ (see \eqref{equz}). Thus, the Hopf locus is parametrized by
\begin{equation}
\left\lbrace
    \begin{array}{l}
    p=p_{H}(z)=-\dfrac{A_\psi(z)}{B_\psi(z)},\vspace{0.2cm}  \\
    u=u_{H}(z)=\dfrac{p_H(z)}{k+z}+\dfrac{s}{m+z}-\dfrac{w}{z},\vspace{0.2cm}  \\
    \end{array}
    \label{paramuphopf}
\right.
\end{equation}
which is defined for $B_\psi(z) \neq 0$. Since $j_{31}<0$ and $(1-z)/(v+z-1)>0$ (see \eqref{eqxc}), we conclude that $B_\psi(z) = 0$ if and only if $sz/(m+z)= j_{33}$, which is equivalent to the third degree polynomial equation
\begin{equation}
    p_B(z)=m \left(v (c+2 z-1)+(z-1)^2\right)+c v (s+z)+z \left(v (s+2 z-1)+(z-1)^2\right)=0.
    \label{eqpoldenb}
\end{equation}
Excluding the possible roots of \eqref{eqpoldenb}, \eqref{paramuphopf} is well-defined for $z\in I_H=\{z\in(0,1);\; p_B(z)\neq 0\}$. If we fix $u$ and adopt $\zeta=p$ as the bifurcation parameter of Theorem \eqref{thm:hopf}, this also guarantees the last condition in \eqref{conditionsLiu}, since ${{d\psi}}/{{d}p}=B_\psi\neq 0$. To guarantee that that $p_H(z)$ and $u_H(z)$ do not become negative, we conclude that $I_H$ needs to be restricted to
$$
I_H=\{z\in(0,1);\;p_B(z)\neq 0,\; -\dfrac{A_\psi(z)}{B_\psi(z)}>(k+z)\left(\frac{w}{z}-\frac{s}{m+z}\right) \},
$$
since $u_H(z)>0$ if and only if
\[
p_H(z)>(k+z)\left(\frac{w}{z}-\frac{s}{m+z}\right)>0.
\]
Finally, we note that the Hopf locus is contained in the image of \eqref{paramuphopf}, but is not necessarily equal to it. This is because \eqref{paramuphopf} gives all parameter pairs for which $\psi=d_2d_1-d_0=0$, and this condition may also include a neutral saddle. Indeed, if a polynomial $c_0+c_1\lambda+c_2 \lambda^2+\lambda^3$ has real roots $l_1,l_2,l_3$ with $l_2=-l_1$, then $c_2c_1-c_0=0$, with $c_2c_0<0$. Thus, $\psi=d_2d_1-d_0=0$ is a condition also for a neutral saddle. To differ between the two cases, we note that in a Hopf point, the coefficients satisfy $d_2d_0>0$. Since $d_2>0$ under the hypothesis of Lemma \ref{lemmaB}, the interval for $z$ which gives the Hopf locus is $\{z\in I_H;\ d_0>0\}$. A boundary point for this is $z_{*}$ such that $d_0=0$, which implies, from \eqref{eq:d0Jc}, in $\phi(z)=0$, corresponding to the a point at the saddle-node curve. In summary, if the curve \eqref{paramuphopf} intersects the saddle-node curve, only one part of it corresponds to the Hopf locus for $E_C$, while the other corresponds to locus of neutral saddle for $E_S$. The possibility of such intersection also suggests the existence of a Bogdanov-Takens bifurcation, as we shall see below. For the parameter values used in Figures \ref{fig:regsefs2} and \ref{fig:regsefs3}, the interval $I_H$ is $(0.15,0.172041)$, with the left boundary corresponding to $d_0=0$ and the right to $u_H=0$.

With the above results, we have proved the following.

\begin{theorem}
    \label{thmhop}
     Assume the conditions of Lemma \ref{lemmaB} hold, denote by $\lambda_1$ the real negative eigenvalue of $J_C$ given by Theorem \ref{thmstb}, and by $\lambda_2$ and $\lambda_3$ the other two eigenvalues of $J_C$. Then, there exists an open interval $I_H$ and a curve $C_H$ parameterized by
    \[
    z \mapsto (p_H(z),u_H(z))\in C_H,\ z\in I_H,
    \]
    given in \eqref{paramuphopf}, such that $\lambda_2$ and $\lambda_3$ form a pair of complex eigenvalues with zero real part if and only if $(p,u)=(p_H(z),u_H(z))$ for some $z \in I_H$. Moreover, $C_H$ divides each of regions $R_1$ and $R_2$ into two subregions $R_i^s$ and  $R_i^u$, $i=1,2$, such that $\lambda_2$ and $\lambda_3$ have negative real part in $R_i^s$ and positive real part in $R_i^u$. For $(p,u)$ close to $(p_H(z),u_H(z))$ in one of the sides $R_i^s$ or $R_i^u$, there exists a small-amplitude periodic solution $O_C$ near $E_C$.
\end{theorem}

The criterion by Liu does not inform whether the Hopf bifurcation is subcritical, supercritical or degenerate. Thus, we cannot infer the stability of the periodic solution $O_C$ and whether it lies in $R_i^s$ or $R_i^u$. In all of our simulations, we found that it is always exists in $R_i^u$ and is stable, implying that the bifurcation is supercritical, so that the attractor changes from the equilibrium point $E_C$ to the limit cycle $O_C$ as $(p,u)$ goes from $R_i^s$ to $R_i^u$. We shall see that these findings are valid under the assumptions of Theorem \ref{theo:BT}, which guarantees the existence of a Boganov-Takens bifurcation, which in turn allows us to infer the exact type of the Hopf bifurcation.

\subsection{Bogdanov-Takens Bifurcation}

The possibility that the Hopf curve $\cal{H}$ intersects the saddle-node curve $u=u_{SN}$ suggests the existence of a Bogdanov-Takens point at the intersection point, where the Jacobian matrix has a pair of purely imaginary eigenvalues and a zero eigenvalue.

In order to provide an analytical proof, we first establish some simplifying relations between the model parameters, which allow us to easily calculate the quantities necessary to prove the existence of a Bogdanov-Takens bifurcation. We start by noting that setting $m=k$ allows to obtain simple expressions for the roots of $\Gamma(z)$. Indeed, for $m=k$, the negative root of $\Gamma(z)$ is $z_1=-k$ (see \eqref{polz}), while the other two are
\[
z_{2,3}=\dfrac{p+s-w-k u\pm \sqrt{(-k u+p+s-w)^2-4 k u w}}{2 u}.
\]
This also simplifies the expression for $u_{SN}$, obtained by setting the discriminant in the last equation to zero,
\[
u_{SN}(p)=\frac{p+s+w-2 \sqrt{w (p+s)}}{k}.
\]
To eliminate the square root, we set $p+s=4w$, leading to $p=4w-s$ and then $u=w/k$ along the saddle-node curve, while the roots become
\[
z_{2,3}=k.
\]
With these choices, and substituting $x_i=x_C(z_i)$, the Jacobian matrix \eqref{jacobian} along the saddle-node curve becomes
\[
J_i=
\left(
\begin{array}{ccc}
 -\frac{s}{2} & k q & -\frac{(k-1) s (c+k)}{4 k (k+v-1)} \\
 s & -2 k q & \frac{(k-1) s (c+k)}{2 k (k+v-1)} \\
 -\frac{k (k+v-1)^2}{v (c+k)} & 0 & -\frac{k \left(v (c+2 k-1)+(k-1)^2\right)}{v (c+k)} \\
\end{array}
\right)
\]
The Jacobian can be further simplified by setting $q=s$ and $k=c$, leading to
\[
J_i=
\left(
\begin{array}{ccc}
 -\frac{s}{2} & c s & -\frac{(c-1) s}{2 (c+v-1)} \\
 s & -2 c s & \frac{(c-1) s}{c+v-1} \\
 -\frac{(c+v-1)^2}{2 v} & 0 & \frac{-c (c+3 v-2)+v-1}{2 v} \\
\end{array}
\right)
\]
One can easily see that $d_0=\det(J_i)=0$, confirming that we're along the saddle-node curve. A Bogdanov-Takens point along the saddle-node curve is further characterized by $d_2d_0-d_1=0$, which implies $d_1=0$. Calculating $d_1$ as the sum of the second-order principal minors of $J_i$ we obtain
\[
d_1=\frac{c s (2 c (c+3 v-2)-v+2)}{2 v}.
\]
Solving $d_1=0$ for $v$ we obtain
\[
v=\frac{2 (c-1)^2}{1-6 c}
\]
In summary, with the above choices for $m,p,u,k,q$ and $v$ we obtain a Bogdanov-Takens point.

In order to formalize our findings and provide a rigorous result, we now recall some aspects of the Bogdanov-Takens bifurcation theory, which can be found in \cite{Kuznetsov} and \cite{Kuznetsov1} and in the references therein.

\smallskip

Consider system \eqref{eq:ED2}
where now $\zeta=(\kappa,\chi) \in V$ is a vector with two bifurcation parameters. Denote the characteristic polynomial of $J({\mathbf{x}},\zeta)$ as in \eqref{eq:charac_polynomial_PE}.
%
%
%
%
A Bogdanov-Takens point of \eqref{eq:ED2} is a pair $(\mathbf{x}_0,\zeta_0) \in U\times V$ that satisfies
\begin{equation*}
f(\mathbf{x}_0,\zeta_0) = 0,\quad d_0(\mathbf{x}_0,\zeta_0) = 0,\quad d_1(\mathbf{x}_0,\zeta_0) = 0,\quad d_2(\mathbf{x}_0,\zeta_0) \neq 0.
\end{equation*}
Equivalently, a Bogdanov-Takens point is an equilibrium point $\mathbf{x}_0 \in U$ of \eqref{eq:ED2} such that for $\zeta_0 \in V$, the matrix $J(\mathbf{x}_0,\zeta_0)$ is similar to the matrix
\[
\begin{pmatrix}
0 & 1 & 0 \vspace{0.2cm}\\
0 & 0 & 0 \vspace{0.2cm}\\
0 & 0 & \lambda_3
\end{pmatrix}
,
\]
with $\lambda_3 \neq 0$.

\smallskip

A Bogdanov-Takens point $(\mathbf{x}_0,\zeta_0) \in U \times V$ of \eqref{eq:ED2} is transversal if the map
\[
(\mathbf{x}_0,\zeta_0) \longmapsto \left(f(\mathbf{x},\zeta),d_1(\mathbf{x},\zeta),d_0(\mathbf{x},\zeta)\right)
\]
is regular at the point $(\mathbf{x}_0,\zeta_0)$, that is, the regularity condition $R(\mathbf{x}_0,\zeta_0)\neq 0$, where
\begin{equation*}\label{eq:R}
R(\mathbf{x}_0,\zeta_0)=\det 
\begingroup 
\setlength\arraycolsep{2pt}
\begin{pmatrix}
\dfrac{\partial}{\partial x}f_1(\mathbf{x}_0,\zeta_0) & \dfrac{\partial}{\partial y}f_1(\mathbf{x}_0,\zeta_0) & \dfrac{\partial}{\partial z}f_1(\mathbf{x}_0,\zeta_0) & \dfrac{\partial}{\partial \kappa}f_1(\mathbf{x}_0,\zeta_0) & \dfrac{\partial}{\partial \chi}f_1(\mathbf{x}_0,\zeta_0) \vspace{0.2cm}\\
\dfrac{\partial}{\partial x}f_2(\mathbf{x}_0,\zeta_0) & \dfrac{\partial}{\partial y}f_2(\mathbf{x}_0,\zeta_0) & \dfrac{\partial}{\partial z}f_2(\mathbf{x}_0,\zeta_0) & \dfrac{\partial}{\partial \kappa}f_2(\mathbf{x}_0,\zeta_0) & \dfrac{\partial}{\partial \chi}f_2(\mathbf{x}_0,\zeta_0) \vspace{0.2cm}\\
\dfrac{\partial}{\partial x}f_3(\mathbf{x}_0,\zeta_0) & \dfrac{\partial}{\partial y}f_3(\mathbf{x}_0,\zeta_0) & \dfrac{\partial}{\partial z}f_3(\mathbf{x}_0,\zeta_0) & \dfrac{\partial}{\partial \kappa}f_3(\mathbf{x}_0,\zeta_0) & \dfrac{\partial}{\partial \chi}f_3(\mathbf{x}_0,\zeta_0) \vspace{0.2cm}\\
\dfrac{\partial}{\partial x}d_0(\mathbf{x}_0,\zeta_0) & \dfrac{\partial}{\partial y}d_0(\mathbf{x}_0,\zeta_0) & \dfrac{\partial}{\partial z}d_0(\mathbf{x}_0,\zeta_0) & \dfrac{\partial}{\partial \kappa}d_0(\mathbf{x}_0,\zeta_0) & \dfrac{\partial}{\partial \chi}d_0(\mathbf{x}_0,\zeta_0) \vspace{0.2cm}\\
\dfrac{\partial}{\partial x}d_1(\mathbf{x}_0,\zeta_0) & \dfrac{\partial}{\partial y}d_1(\mathbf{x}_0,\zeta_0) & \dfrac{\partial}{\partial z}d_1(\mathbf{x}_0,\zeta_0) & \dfrac{\partial}{\partial \kappa}d_1(\mathbf{x}_0,\zeta_0) & \dfrac{\partial}{\partial \chi}d_1(\mathbf{x}_0,\zeta_0)
\end{pmatrix}
\endgroup
.
\end{equation*}

Since $(\mathbf{x}_0,\zeta_0)$ is a Bogdanov-Takens point, there are eigenvectors $Q_0,P_1 \in \mathbb{R}^{3}$ associated with the zero eigenvalues of $M=J(\mathbf{x}_0, \zeta_0)$ and satisfying $MQ_0=0$ and $M^{T}P_1=0$, where $M^{T}$ stands for the transpose of the matrix $M$. There are also generalized eigenvectors $Q_1,P_0 \in \mathbb{R}^{3}$ such that $MQ_1=Q_0$ and $M^{T}P_0=P_1$. These eigenvectors can always be chosen satisfying the conditions
\begin{equation}\label{eq:N}
\left \langle Q_0,P_1 \right\rangle=\left \langle Q_1,P_0 \right\rangle=0,\quad \left \langle Q_0,P_0 \right\rangle=\left \langle Q_1,P_1 \right\rangle=1,
\end{equation}
where $\left \langle \cdot,\cdot \right\rangle$ is the usual inner product in $\mathbb{R}^n$.

Using the translation $\bar{\mathbf{x}}=\mathbf{x}-\mathbf{x}_0$ and for $\zeta=\zeta_0$ define the function $F$ by $F\left(\mathbf{x}\right)=f\left(\mathbf{x}+\mathbf{x}_0,\zeta_0\right)$ omitting the bar. The function $F$ has Taylor series expansion around $\mathbf{x}=0$ up to terms of order $2$ of the form
\[
F(\mathbf{x})=M\mathbf{x}+\dfrac{1}{2}B(\mathbf{x},\mathbf{x})+O(\|\mathbf{x}\|^{3}),
\]
with $M=J(\mathbf{x}_0,\zeta_0)$ as before, $O\left(\|\mathbf{x}\|^{3}\right)$ a function with Taylor series expansion starting with at least cubic terms and
\[
B(\mathbf{y},\mathbf{z})=\left. \dfrac{\partial^{2}}{\partial \vartheta_1\partial \vartheta_2}F\left(\vartheta_1\mathbf{y}+\vartheta_2\mathbf{z}\right)\right|_{(\vartheta_1,\vartheta_2)=(0,0)},
\]
where $\mathbf{y}=(y_1,y_2,y_3)$ e $\mathbf{z}=(z_1,z_2,z_3)$. The symmetric multilinear function $B$ has components
\begin{equation}\label{eq:CFB}
B_{i}(\mathbf{y},\mathbf{z})=\sum_{j,k=1}^{3}\left.\dfrac{\partial^{2}}{\partial \omega_j\partial \omega_k}F\left(\mathbf{\omega}\right)\right|_{\omega=0}y_jz_k,\quad i=1,2,3,
\end{equation}
where $\mathbf{\omega}=(\omega_1,\omega_2,\omega_3)$.

\smallskip

Based on the above notations and definitions, we can state the Bogdanov-Takens Bifurcation Theorem \cite{Kuznetsov,Kuznetsov1}.

\begin{theorem}\label{theo:BT}
Consider system \eqref{eq:ED2} and let $(\mathbf{x}_0,\zeta_0) \in U\times V$ be a point satisfying:
\begin{description}
\item[a.] $(\mathbf{x}_0,\zeta_0)$ is a transversal Bogdanov-Takens point, where $\zeta_0=(\kappa_0,\chi_0)$;
\item[b.] $a_{BT}=\dfrac{1}{2}\left \langle P_1,B\left(Q_0,Q_0\right)\right \rangle \neq 0$;
\item[c.] $b_{BT}=\left \langle P_0,B\left(Q_0,Q_0\right)\right \rangle+\left \langle P_1,B\left(Q_0,Q_1\right)\right \rangle \neq 0$.
\end{description}
Then system \eqref{eq:ED2} undergoes a Bogdanov-Takens bifurcation at $(\mathbf{x}_0,\zeta_0)$. The dynamics of \eqref{eq:ED2} restricted to the two-dimensional central manifold is locally topologically equivalent to the dynamics of the normal form
\[
\left\{
{\setlength\arraycolsep{2pt}
\begin{array}{rcl}
x' & = & y, \vspace{0.2cm}\\
y' & = & \beta_1+\beta_2 x+a_{BT} x^2+b_{BT} xy,
\end{array}}
\right.
\]
where $\beta_1$ and $\beta_2$ are unfolding parameters of $\kappa-\kappa_0$ and $ \chi-\chi_0$, respectively, and $x,y \in \mathbb{R}$.
\end{theorem}

\smallskip

The next result is an application of Theorem \ref{theo:BT} to system \eqref{sys2}.

\begin{theorem}
Consider the following choice of parameters
\begin{equation}\label{eq:EP}
w>0,\:\:\: 0<c< \dfrac{1}{6},\:\:\: m=k=c, \:\:\: 0<s<\dfrac{1}{2},\:\:\: s\neq \dfrac{1}{2(4c+1)},\:\:\: q=s,\:\:\: v=\dfrac{2 (c-1)^2}{1-6 c}.
\end{equation}
Then system \eqref{sys2} undergoes a Bogdanov-Takens bifurcation at $(x_0,y_0,z_0,p_0,u_0)$ with
\[
x_0 = \dfrac{2 (1-6 c) c}{4 c+1},\quad y_0 = \dfrac{1-6 c}{4 c+1},\quad z_0 = c,\quad p_0 = 4 w-s,\quad u_0 = \dfrac{w}{c},
\]
where $x_0$, $y_0$ and $z_0$ are the coordinates of the equilibrium point and $(p_0,u_0)$ is the bifurcation parameter vector.
\label{thmbt}
\end{theorem}

\begin{proof}
Keeping the notation used in \eqref{eq:ED2}, consider the vector field $f=f(\mathbf{x},\zeta)$ associated with system \eqref{sys2} and given by
\begin{equation}\label{eq:fCARTM}
{\setlength\arraycolsep{2pt}
\begin{array}{rcl}
f(\mathbf{x},\zeta) & = & (f_1(\mathbf{x},\zeta),f_2(\mathbf{x},\zeta),f_3(\mathbf{x},\zeta))\vspace{0.2cm}\\
& = & \left(p \dfrac{x z}{k+z}-w x+q z y-u z x,s x-q z y-m q y,z(1-z)-v \dfrac{x z}{c+x+z}\right),
\end{array}}
\end{equation}
where the parameters $w$, $c$, $m$, $k$, $s$, $q$ and $v$ are chosen as in \eqref{eq:EP} and $\zeta=(p,u)$ is the parameter vector. 

\smallskip

\noindent When $\zeta=\zeta_0=(p_0,u_0)$, where $p_0 = 4 w-s$ and $u_0 = w/c$, system \eqref{sys2} has three equilibrium points
\[
\mathbf{x}_0=E_c=\left(\dfrac{2 (1-6 c) c}{4 c+1},\dfrac{1-6 c}{4 c+1},c\right), \quad \mathbf{x}_1=E_0=(0,0,0),\quad  \mathbf{x}_2=E_T=(0,0,1).
\]
The Jacobian matrix of \eqref{eq:fCARTM} evaluated at the equilibrium point of interest $\mathbf{x}_0$, namely,
\[
J(\mathbf{x}_0,\zeta_0)=\left(
\begin{array}{ccc}
 -\dfrac{s}{2} & c s & \dfrac{s-6 c s}{8 c+2} \vspace{0.2cm}\\
 s & -2 c s & \dfrac{(6 c-1) s}{4 c+1} \vspace{0.2cm}\\
 \dfrac{(4 c+1)^2}{24 c-4} & 0 & \dfrac{1}{4} 
\end{array}
\right),
\]
has a double zero eigenvalue $\lambda_1=\lambda_2=0$ and an additional eigenvalue
\begin{equation}\label{eq:lambda3}
\lambda_3=-\dfrac{1}{4} (1-2 (4 c+1) s).
\end{equation}

\noindent Therefore $(\mathbf{x}_0,\zeta_0)$ is a transversal Bogdanov-Takens point of system \eqref{sys2} since the regularity condition is equivalent to 
\[
R(\mathbf{x}_0,\zeta_0)=\dfrac{c^4 (4 c+1)^2 (c (16 c+3)+1) s^3}{16(c-1)}
\]
and the inequalities for $c$ and $s$ in \eqref{eq:EP} imply that both $\lambda_3$ and $R(\mathbf{x}_0,\zeta_0)$ are nonzero.

\smallskip

\noindent The eigenvectors $Q_0,Q_1,P_0,P_1 \in \mathbb{R}^3$ associated with the double zero eigenvalue satisfying $MQ_0=0$, $MQ_1=Q_0$, $M^{T}P_1=0$ and $M^{T}P_0=P_1$ and also the conditions \eqref{eq:N} are given by
\begin{equation*}
{\setlength\arraycolsep{2pt}
\begin{array}{rcl}
Q_0 & = & \left(\dfrac{1-6 c}{(4c+1)^2},\dfrac{2 (6c-1)}{(4 c+1)^2},1\right),\vspace{0.2cm}\\
Q_1 & = & \left(\dfrac{8 (6 c-1) s}{(4 c+1) (2 (4c+1)s-1)},\dfrac{(6 c-1) (4 (4 c+1) (s-1) s+1)}{c(4 c+1)^2 s (2 (4c+1)s-1)},-\dfrac{4}{2 (4c+1)s-1}\right),\vspace{0.2cm}\\
P_0 & = & \left(\dfrac{(4 c+1)^2}{(6 c-1) (2 (4c+1)s-1)},0,\dfrac{1}{2 (4c+1)s-1}+1\right),\vspace{0.2cm}\\
P_1 & = & \left(\dfrac{2 c (4 c+1)^2 s}{(6 c-1) (2 (4c+1)s-1)},\dfrac{c (4 c+1)^2 s}{(6 c-1) (2 (4c+1)s-1)},0\right).
\end{array}}
\end{equation*}

\smallskip

From \eqref{eq:fCARTM} and \eqref{eq:CFB} we obtain
\begin{equation*}
B(\mathbf{y},\mathbf{z})=\left(B_1(\mathbf{y},\mathbf{z}),B_2(\mathbf{y},\mathbf{z}),B_3(\mathbf{y},\mathbf{z})\right),
\end{equation*}
where
\[
{\setlength\arraycolsep{2pt}
\begin{array}{rcl}
B_1(\mathbf{y},\mathbf{z}) & = & \dfrac{-2 (6 c-1) y_3z_3(s-4w)+(4 c+1) s z_3 (4 c y_2-y_1)+(4 c+1) s y_3 (4 c z_2-z_1)}{4 c (4 c+1)}, \vspace{0.2cm}\\
B_2(\mathbf{y},\mathbf{z}) & = & -s (y_2 z_3+y_3 z_2), \vspace{0.2cm}\\
B_3(\mathbf{y},\mathbf{z}) & = & \dfrac{n_3(\mathbf{y},\mathbf{z})}{8 (c-1)c (6 c-1)},
\end{array}}
\]
with
\[
n_3(\mathbf{y},\mathbf{z})=(4 c+1)^2 y_1 (4 c z_1+(7 c-2) z_3+z_1)+(7c-2) (4 c+1)^2 y_3z_1+(2 c (4 c (12 c+7)-15)+3) y_3z_3.
\]

\smallskip

It follows that the coefficients 
\[
{\setlength\arraycolsep{2pt}
\begin{array}{rcl}
a_{BT} & = & \dfrac{2 (4 c+1) s w}{2 (4c+1)s-1},\vspace{0.2cm}\\
b_{BT} & = & \dfrac{128 c^4 s^2+8 c^3 s (7 s-2)+c^2 \left(14 s^2+s (16 w-3)-8 w\right)+c (2 s-1) (s-6 w)+(2-4 s) w}{(c-1) c (2 (4c+1)s-1)^2},
\end{array}}
\]
calculated as in Theorem \ref{theo:BT} are nonzero with the parameter choices \eqref{eq:EP}. In fact, the claim is easily verified for the coefficient $a_{BT}$. However, for the coefficient $b_{BT}$ there is only one value of $w$, namely,
\begin{equation}\label{eq:wc}
w_c=\dfrac{c \left(16 c^2+3 c+1\right) s (8 c s+2
   s-1)}{2 (1-c) (4 c+1) (2 s-1)}
\end{equation}
which makes it zero. But for $0<c<1/6$, on the one hand $0<w_c<s$ if $0<s<1/(2(4c+1))$ contrary to the condition $w>s$ and on the other hand $w_c<0$ if $1/(2(4c+1))<s<1/2$. So the theorem is proved.
\end{proof}

Figure \ref{fig:regsefs3} shows the bifurcation diagram of \eqref{sys2} for the numerical choice of parameters $v=289/20,\ w=1/2,\ m=k=c=3/20,\ q=s=9/20$, 
which satisfies \eqref{eq:EP}. In this case a Bogdanov-Takens bifurcation occurs when $\zeta_0=(p_0,\zeta_0)=\left(31/20,10/3\right)$ and the coordinates of the equilibrium point under these conditions are $x_0=3/160$, $y_0=1/16$ and $z_0=3/20$.

In particular, Theorem \ref{thmbt} implies that, in the vicinity of the bifurcation parameters, the Hopf bifurcation found by Theorem \ref{thmhop} is supercritical, providing an analytical proof for the numerical results showing stable limit cycles around the tumor control equilibrium.

The parameter choices made in \eqref{eq:EP} for the occurrence of a Bogdanov-Takens bifurcation in the CAR-T model \eqref{sys2} are by no means the most general ones. However, these choices avoid a numerical analysis and show that degenerate cases can occur in this model. For example, there is a value $s>0$ for which $\lambda_3$ given in \eqref{eq:lambda3} is zero, possibly giving rise to a triple zero eigenvalue bifurcation. There are also choices for the positive parameters $c$ and $s$ that make $w_c$ given in \eqref{eq:wc} greater than $s$, and therefore the coefficient $b$ can change sign at $w=w_c$ providing either a new bifurcation diagram when $w \neq w_c$ or a degenerate Bogdanov-Takens bifurcation when $w=w_c$. However, these aforementioned cases will not be studied here nor will other bifurcations such as Bautin (or degenerate Hopf bifurcation), cusp and zero-Hopf bifurcation, which seem to exist based on numerical evidence.

\section{Discussion}
\label{sec:conclusion}

The relatively recent advent of CAR-T cell therapy has opened new avenues for mathematical modeling. The need to better understand the underlying mechanisms, quantify clinical and pharmacokinetic variables, and the availability of robust data has spurred the development of various models capturing several aspects of CAR-T cell therapy \cite{stein2019tisagenlecleucel,liu2021model,paixao2022modeling,santurio2024mechanisms,chaudhury2020chimeric,reviewperez,altrock2021,perez-garcia,rockne,Kareva2023Cytokine,mueller2021early,singh2021bench,salem2023development,barros2021cart,serrano2024understanding,perez2021car,santurio2024mathematical}. Some studies have developed approaches to predict long-term therapy outcomes based on short-term responses, but this remains challenging \cite{paixao2022modeling,perez-garcia,mueller2021early,awasthi2020tisagenlecleucel,hay2017chimeric,vercellino2020predictive,kirouac2023deconvolution}. Few studies have conducted a deep qualitative analysis of the nonlinear dynamics in CAR-T cell therapy, which can provide insights into long-term outcomes \cite{barros2021cart,serrano2024understanding,perez2021car}. However, these studies often focused on simplified models or lacked integration with clinical data. This study addresses this gap by rigorously analyzing a complex model that describes real patient data over both short and long-term responses. While the short-term behavior of our model was studied in \cite{paixao2022modeling}, providing insights into the multiphasic kinetics observed in the first weeks after CAR-T cell injection, here we explore the long-term behavior by examining the various bifurcations that occur in the permanent system associated with our model.

Our findings on this nonlinear dynamics that unfold in CAR-T cell therapy provide important biological insights and suggest that complex tumor-immune interactions can lead to nonintuitive outcomes. Namely, the nontrivial dependence on the initial tumor burden required for tumor control in the bistable regime, the role of increased CAR-T cell expansion in driving oscillatory rather than damped tumor control via a Hopf bifurcation, and the influence of tumor-induced immunosuppression in transforming oscillatory tumor control into transient control followed by escape via a Bogdanov-Takens bifurcation. Similar phenomena have been reported in earlier work on tumor-immune interactions. For example, tumor escape for a small initial number of tumor cells, described by Kuznetsov as the "sneaking through" phenomenon \cite{kuznetsov1994nonlinear}, where small numbers of tumor cells evade immune response while larger numbers are eliminated. Delayed tumor escape, where the trajectory initially approaches the unstable tumor control equilibrium before escaping, was also described by Delgado et al. \cite{delgado2020bautin} in their bifurcation analysis of the minimal model of immunoediting by DeLisi and Rescigno \cite{delisi1977immune}.

Our model contains two nonlinear terms that appear to be responsible for most of its complex dynamics, namely the expansion of CAR-T cells and the killing of tumor cells. To better understand their role, it is instructive to compare this model with the three-dimensional ODE model for CAR-T cell therapy in preclinical studies analyzed by one of us in \cite{barros2021cart}. There are two primary structural differences between that model and model \eqref{system1red}. Using the notation of \eqref{system1red}, the first difference is in the expansion of effector CAR-T cells. In \cite{barros2021cart}, this is described by an exponential growth term $r_{\min} C_T$. In contrast, our model incorporates the tumor cell presence by describing expansion as $r_{\min} C_T T/(A+T)$, meaning CAR-T cells must bind to antigen-expressing tumor cells to expand. This nonlinear response aligns with classical T cell biology \cite{antia2005role} and established models of T cell expansion dependent on antigen recognition \cite{de2001recruitment,antia2003models,de2013quantifying}. The second difference lies in the tumor cell killing mechanism. While \cite{barros2021cart} uses a mass action term $-\gamma C_T T$, our model uses $-\gamma C_T T/(a+C_T+dT)$, a modified Holling type II functional response with an additional $C_T$ term, also known as a Beddington-DeAngelis response \cite{haque2011detailed}. This modification accounts for the saturation of tumor cell killing due to the limited number of receptors on the surface of CAR T cells and the resulting competition among CAR T cells to bind to tumor cells. Therefore, the model in \cite{barros2021cart} can be seen as a linearization or limit case of our system \eqref{system1red}. It is worth to note that the bifurcation diagram in \cite{barros2021cart} is structurally similar to the simplest case here (Figure \ref{fig:regsefs}), featuring regions of tumor escape, control, and bistability, but no limit cycles. Thus, we conclude that the refined formulation of these terms is system \eqref{system1red} leads to the more complex nonlinear dynamics with Hopf and Bogdanov-Takens bifurcations, as shown in Figures \ref{fig:regsefs2} and \ref{fig:regsefs3}.

This complexity comes at the cost of interesting mathematical challenges. The third-degree polynomial equation for the nontrivial equilibria presents a challenge for analyzing their existence and stability. We addressed this by parameterizing the bifurcation curves and manipulating the coefficients of the Jacobian matrix. To explore the loss of stability at the tumor control equilibrium point and potential Hopf bifurcation, we used Liu’s criterion to characterize the Hopf bifurcation manifold. By parameterizing it through solving a two-dimensional algebraic system, we derived an analytical expression for the Hopf locus. Additionally, by imposing conditions on model parameters and eliminating parameters $m, k, q, v$, we simplified expressions to provide an analytical proof of the Bogdanov-Takens bifurcation. This combination of techniques enabled a balance between a general description of the parameter space and a detailed analysis of specific subsets. Indeed, while Liu’s approach applied to the full parameter space without simplifications but did not indicate the nature of the Hopf bifurcation, the Bogdanov-Takens bifurcation analysis under simplified conditions inferred that the Hopf bifurcation is supercritical.

In summary, this work provides critical insights into the understanding of CAR-T cell therapy by dissecting the nonlinear dynamics of a validated model arising from the complex interplay between CAR-T cells and tumor cells. It also illustrates how combining different mathematical approaches can effectively address nonlinear systems.

\section{Acknowledgments}
A. C. Fassoni was supported by Alexander von Humboldt Foundation and Coordena\c c\~ao de Aperfei\c coamento de Pessoal de N\'ivel Superior - Brasil (CAPES) - Finance Code 001, and partially supported by FAPEMIG RED-00133-21. D. C. Braga was partially supported by FAPEMIG APQ-01158-17 and RED-00133-21.


\begin{thebibliography}{10}

\bibitem{grupp2013chimeric}
Stephan~A Grupp, Michael Kalos, David Barrett, Richard Aplenc, David~L Porter,
  Susan~R Rheingold, David~T Teachey, Anne Chew, Bernd Hauck, J~Fraser Wright,
  et~al.
\newblock Chimeric antigen receptor--modified t cells for acute lymphoid
  leukemia.
\newblock {\em New England Journal of Medicine}, 368(16):1509--1518, 2013.

\bibitem{stein2019tisagenlecleucel}
Andrew~M Stein, Stephan~A Grupp, John~E Levine, Theodore~W Laetsch, Michael~A
  Pulsipher, Michael~W Boyer, Keith~J August, Bruce~L Levine, Lori Tomassian,
  Sweta Shah, et~al.
\newblock Tisagenlecleucel model-based cellular kinetic analysis of chimeric
  antigen receptor--t cells.
\newblock {\em CPT: pharmacometrics \& systems pharmacology}, 8(5):285--295,
  2019.

\bibitem{liu2021model}
Can Liu, Vivaswath~S Ayyar, Xirong Zheng, Wenbo Chen, Songmao Zheng, Hardik
  Mody, Weirong Wang, Donald Heald, Aman~P Singh, and Yanguang Cao.
\newblock Model-based cellular kinetic analysis of chimeric antigen receptor-t
  cells in humans.
\newblock {\em Clinical Pharmacology \& Therapeutics}, 109(3):716--727, 2021.

\bibitem{paixao2022modeling}
Emanuelle~A Paix{\~a}o, Luciana~RC Barros, Artur~C Fassoni, and Regina~C
  Almeida.
\newblock Modeling patient-specific car-t cell dynamics: Multiphasic kinetics
  via phenotypic differentiation.
\newblock {\em Cancers}, 14(22):5576, 2022.

\bibitem{santurio2024mechanisms}
Daniela~Silva Santurio, Emanuelle~A Paix{\~a}o, Luciana~RC Barros, Regina~C
  Almeida, and Artur~C Fassoni.
\newblock Mechanisms of resistance to car-t cell immunotherapy: Insights from a
  mathematical model.
\newblock {\em Applied Mathematical Modelling}, 125:1--15, 2024.

\bibitem{shah2019mechanisms}
Nirali~N Shah and Terry~J Fry.
\newblock Mechanisms of resistance to car t cell therapy.
\newblock {\em Nature reviews Clinical oncology}, 16(6):372--385, 2019.

\bibitem{de2013quantifying}
Rob~J De~Boer and Alan~S Perelson.
\newblock Quantifying t lymphocyte turnover.
\newblock {\em Journal of theoretical biology}, 327:45--87, 2013.

\bibitem{de2005validated}
Lisette~G de~Pillis, Ami~E Radunskaya, and Charles~L Wiseman.
\newblock A validated mathematical model of cell-mediated immune response to
  tumor growth.
\newblock {\em Cancer research}, 65(17):7950--7958, 2005.

\bibitem{hassard1981theory}
Brian~D Hassard, Nicholas~D Kazarinoff, and Yieh-Hei Wan.
\newblock {\em Theory and applications of Hopf bifurcation}, volume~41.
\newblock CUP Archive, 1981.

\bibitem{guckenheimer1986dynamical}
John Guckenheimer.
\newblock Dynamical systems, and bifurcatins of vector fields.
\newblock {\em Appl. Math. Sci. Series}, 42, 1986.

\bibitem{perko2013differential}
Lawrence Perko.
\newblock {\em Differential equations and dynamical systems}, volume~7.
\newblock Springer Science \& Business Media, 2013.

\bibitem{Kuznetsov}
Yuri~A Kuznetsov, Iu~A Kuznetsov, and Y~Kuznetsov.
\newblock {\em Elements of applied bifurcation theory}, volume 112.
\newblock Springer, 1998.

\bibitem{liu1994criterion}
Wei-Min Liu.
\newblock Criterion of hopf bifurcations without using eigenvalues.
\newblock {\em Journal of Mathematical Analysis and Applications},
  182(1):250--256, 1994.

\bibitem{chen2021collective}
Kuan-Wei Chen and Chih-Wen Shih.
\newblock Collective oscillations in coupled-cell systems.
\newblock {\em Bulletin of Mathematical Biology}, 83(6):62, 2021.

\bibitem{shih2022hopf}
Chih-Wen Shih and Chia-Hsin Yang.
\newblock Hopf bifurcation analysis for models on genetic negative feedback
  loops.
\newblock {\em Journal of Mathematical Analysis and Applications},
  516(2):126537, 2022.

\bibitem{liu1987dynamical}
Wei-min Liu, Herbert~W Hethcote, and Simon~A Levin.
\newblock Dynamical behavior of epidemiological models with nonlinear incidence
  rates.
\newblock {\em Journal of mathematical biology}, 25:359--380, 1987.

\bibitem{liu1997nonlinear}
Wei-min Liu.
\newblock Nonlinear oscillations in models of immune responses to persistent
  viruses.
\newblock {\em Theoretical population biology}, 52(3):224--230, 1997.

\bibitem{domijan2009bistability}
Mirela Domijan and Markus Kirkilionis.
\newblock Bistability and oscillations in chemical reaction networks.
\newblock {\em Journal of Mathematical Biology}, 59:467--501, 2009.

\bibitem{Kuznetsov1}
Yu~A Kuznetsov.
\newblock Numerical normalization techniques for all codim 2 bifurcations of
  equilibria in ode's.
\newblock {\em SIAM journal on numerical analysis}, 36(4):1104--1124, 1999.

\bibitem{chaudhury2020chimeric}
Anwesha Chaudhury, Xu~Zhu, Lulu Chu, Ardeshir Goliaei, Carl~H June, Jeffrey~D
  Kearns, and Andrew~M Stein.
\newblock Chimeric antigen receptor t cell therapies: a review of cellular
  kinetic-pharmacodynamic modeling approaches.
\newblock {\em The Journal of Clinical Pharmacology}, 60:S147--S159, 2020.

\bibitem{reviewperez}
Salvador Chulián, Álvaro Martínez-Rubio, María Rosa, and Victor~M
  Pérez-García.
\newblock Mathematical models of leukaemia and its treatment: a review.
\newblock {\em SeMA}, 79:487–488, 2022.

\bibitem{altrock2021}
Altrock~PM Kimmel~GJ, Locke~FL.
\newblock The roles of t cell competition and stochastic extinction events in
  chimeric antigen receptor t cell therapy.
\newblock {\em Proceedings. Biological sciences}, 288:1947, 2021.

\bibitem{perez-garcia}
Álvaro Martínez-Rubio, Salvador Chulián, Manuel Ramírez~Orellana Cristina
  Blázquez~Goñi, Antonio~Pérez Martínez, Cristina~Ferreras Alfonso
  Navarro-Zapata, Victor~M Pérez-García, and María Rosa.
\newblock A mathematical description of the bone marrow dynamics during car
  t-cell therapy in b-cell childhood acute lymphoblastic leukemia.
\newblock {\em International journal of molecular sciences}, 22:6371, 2021.

\bibitem{rockne}
A.~B. Brummer, A.~Xella, R.~Woodall, V.~Adhikarla, H.~Cho, M.~Gutova, C.~E.
  Brown, and R.~C. Rockne.
\newblock Data driven model discovery and interpretation for car t-cell killing
  using sparse identification and latent variables.
\newblock {\em Frontiers in immunology}, 2023.

\bibitem{Kareva2023Cytokine}
I.~Kareva and J.~Gevertz.
\newblock Cytokine storm mitigation for exogenous immune agonists.
\newblock {\em bioRxiv}, 2023.

\bibitem{mueller2021early}
Anna Mueller-Schoell, Nahum Puebla-Osorio, Robin Michelet, Michael~R Green,
  Annette K{\"u}nkele, Wilhelm Huisinga, Paolo Strati, Beth Chasen, Sattva~S
  Neelapu, Cassian Yee, et~al.
\newblock Early survival prediction framework in cd19-specific car-t cell
  immunotherapy using a quantitative systems pharmacology model.
\newblock {\em Cancers}, 13(11):2782, 2021.

\bibitem{singh2021bench}
Aman~P Singh, Wenbo Chen, Xirong Zheng, Hardik Mody, Thomas~J Carpenter, Alice
  Zong, and Donald~L Heald.
\newblock Bench-to-bedside translation of chimeric antigen receptor (car) t
  cells using a multiscale systems pharmacokinetic-pharmacodynamic model: a
  case study with anti-bcma car-t.
\newblock {\em CPT: Pharmacometrics \& Systems Pharmacology}, 10(4):362--376,
  2021.

\bibitem{salem2023development}
Ahmed~M Salem, Ganesh~M Mugundu, and Aman~P Singh.
\newblock Development of a multiscale mechanistic modeling framework
  integrating differential cellular kinetics of car t-cell subsets and
  immunophenotypes in cancer patients.
\newblock {\em CPT: Pharmacometrics \& Systems Pharmacology}, 12(9):1285--1304,
  2023.

\bibitem{barros2021cart}
Luciana~RC Barros, Emanuelle~A Paix{\~a}o, Andrea~MP Valli, Gustavo~T Naozuka,
  Artur~C Fassoni, and Regina~C Almeida.
\newblock Cart math—a mathematical model of car-t immunotherapy in
  preclinical studies of hematological cancers.
\newblock {\em Cancers}, 13(12):2941, 2021.

\bibitem{serrano2024understanding}
Sergio Serrano, Roberto Barrio, {\'A}lvaro Mart{\'\i}nez-Rubio, Juan
  Belmonte-Beitia, and V{\'\i}ctor~M P{\'e}rez-Garc{\'\i}a.
\newblock Understanding the role of b-cells in car t-cell therapy in leukemia
  through a mathematical mode.
\newblock {\em arXiv preprint arXiv:2403.00340}, 2024.

\bibitem{perez2021car}
V{\'\i}ctor~M P{\'e}rez-Garc{\'\i}a, Odelaisy Le{\'o}n-Triana, Mar{\'\i}a Rosa,
  and Antonio Perez-Martinez.
\newblock Car t cells for t-cell leukemias: Insights from mathematical models.
\newblock {\em Communications in Nonlinear Science and Numerical Simulation},
  96:105684, 2021.

\bibitem{santurio2024mathematical}
Daniela~S Santurio, Luciana~RC Barros, Ingmar Glauche, and Artur~C Fassoni.
\newblock Mathematical modeling unveils the timeline of car-t cell therapy and
  macrophage-mediated cytokine release syndrome.
\newblock {\em bioRxiv}, pages 2024--04, 2024.

\bibitem{awasthi2020tisagenlecleucel}
Rakesh Awasthi, Lida Pacaud, Edward Waldron, Constantine~S Tam, Ulrich
  J{\"a}ger, Peter Borchmann, Samantha Jaglowski, Stephen~Ronan Foley, Koen
  Van~Besien, Nina~D Wagner-Johnston, et~al.
\newblock Tisagenlecleucel cellular kinetics, dose, and immunogenicity in
  relation to clinical factors in relapsed/refractory dlbcl.
\newblock {\em Blood advances}, 4(3):560--572, 2020.

\bibitem{hay2017chimeric}
Kevin~A Hay and Cameron~J Turtle.
\newblock Chimeric antigen receptor (car) t cells: lessons learned from
  targeting of cd19 in b-cell malignancies.
\newblock {\em Drugs}, 77:237--245, 2017.

\bibitem{vercellino2020predictive}
Laetitia Vercellino, Roberta Di~Blasi, Salim Kanoun, Benoit Tessoulin, Cedric
  Rossi, Maud D'Aveni-Piney, Lucie Ob{\'e}ric, Caroline Bodet-Milin, Pierre
  Bories, Pierre Olivier, et~al.
\newblock Predictive factors of early progression after car t-cell therapy in
  relapsed/refractory diffuse large b-cell lymphoma.
\newblock {\em Blood advances}, 4(22):5607--5615, 2020.

\bibitem{kirouac2023deconvolution}
Daniel~C Kirouac, Cole Zmurchok, Avisek Deyati, Jordan Sicherman, Chris Bond,
  and Peter~W Zandstra.
\newblock Deconvolution of clinical variance in car-t cell pharmacology and
  response.
\newblock {\em Nature biotechnology}, pages 1--12, 2023.

\bibitem{kuznetsov1994nonlinear}
Vladimir~A Kuznetsov, Iliya~A Makalkin, Mark~A Taylor, and Alan~S Perelson.
\newblock Nonlinear dynamics of immunogenic tumors: parameter estimation and
  global bifurcation analysis.
\newblock {\em Bulletin of mathematical biology}, 56(2):295--321, 1994.

\bibitem{delgado2020bautin}
Joaqu{\'\i}n Delgado, Eymard Hern{\'a}ndez-L{\'o}pez, and Luc{\'\i}a~Ivonne
  Hern{\'a}ndez-Mart{\'\i}nez.
\newblock Bautin bifurcation in a minimal model of immunoediting.
\newblock {\em Discrete \& Continuous Dynamical Systems-Series B}, 25(4), 2020.

\bibitem{delisi1977immune}
Charles DeLisi and Aldo Rescigno.
\newblock Immune surveillance and neoplasia—1 a minimal mathematical model.
\newblock {\em Bulletin of mathematical biology}, 39:201--221, 1977.

\bibitem{antia2005role}
Rustom Antia, Vitaly~V Ganusov, and Rafi Ahmed.
\newblock The role of models in understanding cd8+ t-cell memory.
\newblock {\em Nature Reviews Immunology}, 5(2):101--111, 2005.

\bibitem{de2001recruitment}
Rob~J De~Boer, Mihaela Oprea, Rustom Antia, Kaja Murali-Krishna, Rafi Ahmed,
  and Alan~S Perelson.
\newblock Recruitment times, proliferation, and apoptosis rates during the cd8+
  t-cell response to lymphocytic choriomeningitis virus.
\newblock {\em Journal of virology}, 75(22):10663--10669, 2001.

\bibitem{antia2003models}
Rustom Antia, Carl~T Bergstrom, Sergei~S Pilyugin, Susan~M Kaech, and Rafi
  Ahmed.
\newblock Models of cd8+ responses: 1. what is the antigen-independent
  proliferation program.
\newblock {\em Journal of theoretical biology}, 221(4):585--598, 2003.

\bibitem{haque2011detailed}
Mainul Haque.
\newblock A detailed study of the beddington--deangelis predator--prey model.
\newblock {\em Mathematical Biosciences}, 234(1):1--16, 2011.

\end{thebibliography}

\end{document}